\documentclass[aps,amsmath,prd,groupedaddress,nofootinbib,preprintnumbers,showpacs]{revtex4}

\usepackage{graphicx}

\begin{document}

\preprint{IHES/P/02/10}


\title{Circular orbits of corotating binary black holes: comparison between analytical
and numerical results}


\author{Thibault Damour}
\email[]{damour@ihes.fr}
\affiliation{ Institut des Hautes Etudes Scientifiques, 91440
Bures-sur-Yvette, France}

\author{Eric Gourgoulhon}
\email[]{Eric.Gourgoulhon@obspm.fr}
\affiliation{Laboratoire de l'Univers et de ses Th\'eories,
FRE 2462 du C.N.R.S., Observatoire de Paris, F-92195 Meudon Cedex, France}

\author{Philippe Grandcl\'ement}
\email[]{PGrandclement@northwestern.edu}
\affiliation{Department of Physics and Astronomy, Northwestern University,
Evanston, IL 60208, USA}


\date{2 May 2002}

\begin{abstract}
We compare recent numerical results, obtained within a``helical Killing vector'' (HKV)
approach, on circular orbits of corotating binary black holes to the analytical
predictions made by the effective one body (EOB) method (which has been recently extended
to the case of spinning bodies). On the scale of the differences between the results
obtained by different numerical methods, we find good agreement between numerical data
and analytical predictions for several invariant functions describing the dynamical
properties of circular orbits. This agreement is robust against the post-Newtonian 
accuracy used for the analytical estimates, as well as under choices of resummation
method for the EOB ``effective potential'', and gets better as one uses a higher
post-Newtonian accuracy. These findings open the way to a significant ``merging'' of
analytical and numerical methods, i.e. to matching an EOB-based analytical description
of the (early and late) inspiral, up to the beginning of the plunge, to a numerical
description of the plunge and merger. We illustrate also the ``flexibility'' of the EOB
approach, i.e. the possibility of determining some ``best fit'' values for the analytical
parameters by comparison with numerical data.
\end{abstract}

\pacs{04.30.Db, 04.25.Nx, 04.25.Dm, 04.70.Bw, 97.60.Lf, 97.80.-d}

\maketitle


\section{Introduction}\label{sec1}

Binary black holes are the most promising candidate sources for the LIGO / VIRGO / GEO600 /
TAMA $\ldots$ network of ground based gravitational wave (GW) interferometric detectors 
\cite{LPP97,FH98,BCT98,PZM,DIS00}. Signal to noise ratio estimates suggest that the 
first detections will concern black hole binaries of total mass $\gtrsim 25 \, 
M_{\odot}$, and show that the most ``useful'' part of the gravitational waveform is 
emitted in the last $\sim 5$ orbits of the inspiral, and during the plunge taking place 
after crossing the last stable (circular) orbit (LSO) \cite{DIS00}. This makes it urgent 
to have reliable methods allowing one to model the last orbits of binary black holes.

Recently, Ref.~\cite{BD00} has suggested a new method to tackle the dynamics and GW 
emission from the last orbits of binary black holes. The basic idea of \cite{BD00} was 
to extend, by using suitable resummation methods, the validity of the perturbative 
(``post-Newtonian'') analytical calculations of the equations of motion
\cite{DD,JS98,DJS01,BF00} and GW radiation \cite{BDIWW,B} so as to be able to describe 
not only the last orbits before the LSO, but also the transition to the plunge. The main 
philosophy of \cite{BD00} was to push analytical methods to their limits so as to use 
numerical methods only to describe the plunge, which was estimated (when the masses are 
comparable) to last about 60\% of one orbit. To implement this philosophy, one needs to 
be able to construct {\it numerical} initial data that match the {\it dynamical} initial 
data given by the method of \cite{BD00} at the start of the plunge. This is a non 
trivial task because, up to very recently, there was a significant discrepancy between 
analytical \cite{DIS98,BD00,DJS00,B02} and numerical \cite{cook,PfeifTC00,baumgarte}
estimates of the dynamical characteristics of the orbits near the LSO. 
For instance, the binding
energy at the LSO, $e_{\rm LSO} = E_{\rm LSO} / (m_1 + m_2) - 1$, is analytically
estimated to be around $e_{\rm LSO}^{\rm ana} \simeq - 1.67$\% \cite{DJS00} (third 
post-Newtonian effective-one-body estimate with $\omega_s = 0$ \cite{DJS01}), while two
different numerical estimates \cite{cook,baumgarte} gave $e_{\rm LSO}^{\rm num} \simeq 
-2.3$\%, i.e. a significantly more bound (by 38\%) LSO. The discrepancy is even more 
striking if one considers the LSO orbital period: the analytical estimate is $T_{\rm 
LSO}^{\rm ana} \simeq 71.2 \, (m_1 + m_2)$ \cite{DJS00}, which is twice longer than the 
numerical one $T_{\rm LSO}^{\rm num} \simeq 35 \, (m_1 + m_2)$ \cite{baumgarte}. [By 
contrast, the dispersion among analytical estimates \cite{DIS98,BD00,DJS00,B02}, as well 
as the dispersion among numerical ones \cite{cook,PfeifTC00,baumgarte}
is significantly smaller
than the difference between analytical and numerical results. See \cite{Buona02} for a 
review of analytical methods, and see below for a brief
discussion of our choice of analytical estimates.] This discrepancy  diminishes
the physical relevance of a recent attempt \cite{baker} at fulfilling the proposal of
Ref.~\cite{BD00}, i.e. to start a full numerical calculation of the plunge just after 
the crossing of the LSO.

Very recently, a new numerical approach to the circular orbits of binary black holes has
been set up \cite{GGB1} and implemented \cite{GGB2} (see also Ref.~\cite{Cook02}
for a discussion and an extension to any black hole rotation state).
Contrary to the previous numerical approaches \cite{cook,PfeifTC00,baumgarte}, which were formulated
in terms of the initial value problem of general relativity
and dealt therefore only with the four constraint equations
on a 3-dimensional spacelike hypersurface, the new approach \cite{GGB1}
deals with a full 4-dimensional spacetime
(see Ref.~\cite{Cook02} for the link with York's conformal thin-sandwich
formalism \cite{York99,Cook00}).
The basic assumption is that this
spacetime is endowed with a helical Killing vector \cite{Detwe89}, which amounts to
neglecting the gravitational radiation reaction, i.e. to considering that
the orbits are exactly circular. In addition, Refs.~\cite{GGB1,GGB2} assume,
as a simplifying approximation, that the spatial 3-metric is conformally
flat (see Sec. IV.C of Ref.~\cite{FriedUS02} for a discussion).
The slicing is also chosen to be maximal (i.e. $K=0$).
The spacetime manifold is chosen to have the topology of the real line
times the two-sheeted Misner-Lindquist manifold. The two sheets are assumed
to be isometric (with respect to the full 4-metric). Under the above
assumptions, the problem amounts to solving five of the ten Einstein equations.
This is one more that the previous numerical approaches \cite{cook,PfeifTC00,baumgarte}.
Hereafter we call the new method \cite{GGB1,GGB2} the {\em helical Killing vector
(HKV)} approach, and previous methods \cite{cook,PfeifTC00,baumgarte}
{\em IVP} ones (for {\em Initial Value Problem}).
Note that the HKV approach has also been employed for binary neutron stars (see e.g.
\cite{GourgGTMB01} and references therein).

The first numerical results obtained by the HKV method \cite{GGB2}
for the characteristics
of the LSO indicate a much better agreement with analytical estimates than those
obtained with the IVP method \cite{cook,PfeifTC00,baumgarte}.
However, the comparison, done in \cite{GGB2},
used analytical results valid only for {\it non spinning} black holes, while the
numerical approach of \cite{GGB1} concerns {\it corotating} binary black holes. Moreover,
the comparison made in \cite{GGB2} was limited to the sole LSO, without looking at the
behavior of the other circular orbits. It is known that the presence of even small
additional interactions (especially repulsive ones) can have an important effect on the
dynamical characteristics of the orbits around the LSO. It is therefore a priori
necessary to include spin-orbit effects (which are repulsive in the corotating case) 
when doing the analytical/numerical comparison. Fortunately, a recent analytical work 
\cite{D01} has shown how to generalize the effective one body (EOB) approach of 
\cite{BD00} by including, to lowest order, spin-orbit and spin-spin effects. The purpose 
of the present work is to take into account in detail the effect of the spin 
interactions in corotating binary black holes. Another important feature of this work is 
that we shall compare the predictions of the EOB method \cite{BD00,DJS00,D01} and the 
results of the new numerical approach of \cite{GGB1,GGB2}, not only for the 
characteristics of the LSO (to which most authors limit their considerations), but also 
for all other circular orbits. We shall do that by comparing several physically 
important (and coordinate-invariant) functions, such as the binding energy as a function 
of angular velocity. Note that, in this work, we loosely call ``LSO'', for a corotating
system, the {\it maximal binding} configuration along a corotating sequence.
For HKV sequences, it has been shown that this minimal energy configuration
marks the onset of an orbital instability \cite{FriedUS02}.
The fact that such corotating sequences are (probably) not realized in actual
coalescing systems\footnote{Note, however, the claim of \cite{PriceW01} that the effective
viscosity of black holes might be large enough to trigger a significant angular-momentum
transfer, potentially able to ensure tidal locking at small
separations.},
and that the corotating ``LSO'' signals only a secular instability, is not important
for the present work whose main aim is to show how one can relate the numerical and
analytical descriptions of the same configuration involving spinning black holes.

This paper is organized as follows. Section II discusses the arguments selecting the EOB 
method as the current best analytical method for tackling the last orbits of binary 
black holes and summarizes the EOB approach to spinning black holes. Section III applies 
the EOB formalism to corotating black holes on circular orbits and shows how to compare 
the analytical results to the recent numerical computation of a sequence of corotating 
binary black holes. Section IV discusses the meaning of our results.

\section{Effective one-body approach to spinning binary black holes}\label{sec2}

\subsection{Formulation}

Before summarizing the effective one-body (EOB) formalism, let us recall why we consider 
it as the best current analytical formalism for tackling the dynamics of the last orbits 
of binary black holes. The first point is that the second work in Ref.~\cite{DIS00} has 
shown that it is crucial, if one wishes to keep a good overlap with the expected real 
signals, to generate GW templates which go beyond the ``adiabatic approximation'' and 
which use accurate equations of motion, including radiation reaction, to describe the 
smooth transition (taking place around the LSO) between inspiral motion and plunge. 
However, the most accurate equations of motion \cite{DD,JS98,DJS01,BF00} are a priori 
given as very complicated expressions, containing hundreds of terms. These equations of 
motion are essentially power series in a  formal ``small parameter'' $\epsilon \sim  
GM / c^2r \sim v^2 / c^2$ [``post-Newtonian'' (PN) expansion]. The problem is that the 
``small'' PN expansion parameter $\epsilon$
is not numerically very small near the LSO. It was 
found quite early that the use of straightforward PN-expanded equations of motion (as in 
\cite{LW90}) is unsatisfactory because the PN-expanded equations of motion converge so 
slowly that, at low orders, they can lead to a behavior which is qualitatively 
different from what one expects. In particular, it was shown in \cite{KWW93} that the 
PN-expanded (harmonic-coordinates) equations of motion admit an LSO at the 
second post-Newtonian (2PN) level, but admit no LSO at both the first
post-Newtonian (1PN) and the {\it third post-Newtonian} (3PN) levels. [ ``LSO''
is taken here in the sense of a critical point in the dynamical analysis of circular
orbits.] By contrast, a study of the LSO in terms of the PN-expanded Hamiltonian \cite{WS}
has found that the PN-expanded Hamiltonian (in various coordinates) admits 
no LSO at the 2PN level, but admit LSO's at the 1PN and 3PN levels. This unstable
behavior is a sign of the unreliability of non-resummed PN approaches.

As the recent work \cite{JS98,DJS01,BF00} 
has succeeded in deriving complete 3PN equations of motion, it would be unacceptable to 
``spoil'' the precious information they contain by using them in such an inappropriate 
way. This led Ref.~\cite{KWW93} to propose to improve the situation by using a {\it 
partial resummation} approach. Specifically, they introduced an ``hybrid'' approach in 
which one resums exactly the ``Schwarzschild'' terms in the (relative) equations of 
motion, while continuing to treat the $\nu$-dependent terms (where $\nu \equiv m_1 \,
m_2 / (m_1 + m_2)^2$) as straightforward PN expansions. This ``hybrid'' approach did 
improve the situation to some extent (at least at the 2PN accuracy). However, later work 
showed that the hybrid approach was neither robust, nor consistent. Ref.~\cite{WS} 
showed that the predictions of the hybrid approach were robust neither under using 
Hamiltonian equations of motion instead of Newtonian-like ones, nor under a change of 
coordinate system. Ref.~\cite{DIS98} pointed out the inconsistency of the hybrid method 
by showing that the formal (un-resummed) ``$\nu$-corrections'' represent, in several 
cases, a very large (larger than 100\%) modification of the corresponding resummed 
$\nu$-independent terms. This led to a quest for new resummation approaches
providing full dynamical equations of motion for binary systems, and being robust and 
consistent. The only such method we are aware of is the EOB approach \cite{BD00}. We do 
not consider here  methods such as the $E$- or $e$-methods of \cite{DIS98} and 
the $j$-method of \cite{DJS00} which were devised to investigate the location of the 
LSO. These methods cannot be used beyond the ``adiabatic approximation'' because they
do not provide full dynamical equations of motion able to 
model the transition between inspiral and plunge. Let us note in this respect that the 
work of \cite{BD00} has shown that, in the case of comparable masses, the transition 
between inspiral and plunge is a gradual process in which the precise location of the 
``adiabatic'' LSO is less important than a precise evaluation of radiation damping 
effects. Therefore it is finally not very important to focus on the ``exact'' location 
of the LSO. What is really important is to have a good description of the dynamics of 
the last ten orbits before the plunge. Moreover, even if we were to consider only the 
determination of the LSO, the comparisons made in \cite{DJS00} have shown that the EOB 
method is the most robust in that it is the least sensitive to changes in the 3PN 
coefficients (see Fig.~2 in \cite{DJS00}). We also note in passing that \cite{DIS98} had 
shown that the $e$-method was more robust (both in its dependence on $\nu$ and under 
changes in higher PN coefficients), and more convergent, than the straightforward non
resummed $E$-method (recently used at 3PN in \cite{B02}); see Table I, Table VII and 
Fig.~5 in \cite{DIS98}. The robustness of the EOB method is discussed in 
Refs.~\cite{BD00,DJS00,D01}. We shall give below further examples of the robustness of the
EOB approach.

The basic idea of the EOB method is to map the (complicated) relative dynamics (in the 
center of mass) of a two-body system onto the (drastically simpler) dynamics of an 
``effective'' one-body system. Most of the (physically irrelevant) gauge-related 
complications of the two-body equations of motion get absorbed into the mapping between 
the two problems. The mapping is defined so as to preserve the adiabatic invariants 
(which, at the quantum level, are quantized in units of $\hbar$). The energy mapping 
$E_{\rm eff} = f(E_{\rm real})$ between the real two-body (center of mass) energy 
$E_{\rm real}$ and the effective one-body energy $E_{\rm eff}$
is determined by the matching between the two problems. It 
is remarkably found that the energy map $f$ is extremely simple, and independent of the 
PN accuracy considered (we set $c=1$):
\begin{equation}
\label{eq2.1}
\frac{E_{\rm eff}}{\mu} = \frac{E_{\rm real}^2 - m_1^2 - m_2^2}{2 \, m_1 \, m_2} \, .
\end{equation}
Here, $m_1$ and $m_2$ denote the masses of the two bodies, while $\mu$ denotes the
reduced mass $\mu = m_1 \, m_2 / M$, with $M = m_1 + m_2$ denoting the total mass. Let
us note also the definition of the symmetric mass ratio $\nu \equiv \mu / M = m_1 \, m_2
/ (m_1 + m_2)^2$ ($0 < \nu \leq \frac{1}{4}$). The inverse of the energy map
(\ref{eq2.1}) is
\begin{equation}
\label{eq2.2}
E_{\rm real} = M \, \sqrt{1 + 2 \nu \left( \frac{E_{\rm eff} - \mu}{\mu} \right)} \, .
\end{equation}
The effective energy $E_{\rm eff}$ is given by the effective Hamiltonian $H_{\rm eff}
(\mbox{\boldmath$x$} , \mbox{\boldmath$p$}, \mbox{\boldmath$S$}_1 ,
\mbox{\boldmath$S$}_2)$. Here, $\mbox{\boldmath$x$}$ and $\mbox{\boldmath$p$}$ are the
positions and momenta of the effective one-body (describing the relative motion) and
$\mbox{\boldmath$S$}_1$ and $\mbox{\boldmath$S$}_2$ are the two independent spin vectors
of the two bodies. Ref.~\cite{BD00} considered only the 2PN approximation and the
spin-independent case. The 3PN, spin-independent effective Hamiltonian was determined in
\cite{DJS00}. Ref.~\cite{D01} recently showed how to add the spin interactions in the
EOB framework. The most complete version of the 3PN spin-dependent effective Hamiltonian
is given by Eq.~(2.56) of \cite{D01}. For simplicity, and because spin-spin effects will
turn out to be quite small, we shall consider the less complete, but simpler, effective
Hamiltonian defined by Eq.~(2.26) of \cite{D01} with the effective metric defined by the
``deformed Kerr metric'' of Section IIC there. This Hamiltonian takes into account (at
the lowest PN approximation) the full spin-orbit interaction, and incorporates most of
the spin-spin interaction. [In the case of parallel  spins the ratio between the
spin-spin interaction included in the simplified $H_{\rm eff}$ and the more complete one
$H'_{\rm eff}$ is $\left( \frac{7}{8} \right)^2$.] The Kerr parameter of the effective
Kerr metric describing spin-orbit and spin-spin interactions is given by
\begin{equation}
\label{eq2.3}
M \mbox{\boldmath$a$} \equiv \mbox{\boldmath$S$}_{\rm eff} \equiv \sigma_1 \,
\mbox{\boldmath$S$}_1 + \sigma_2 \, \mbox{\boldmath$S$}_2 \, ,
\end{equation}
where $\sigma_1 \equiv 1 + 3 \, m_2 / (4 \, m_1)$, $\sigma_2 \equiv 1+3 \, m_1 / (4 \, m_2)$.

The EOB approach can deal with the most general case where the spin vectors
$\mbox{\boldmath$S$}_1$, $\mbox{\boldmath$S$}_2$ are arbitrarily oriented. In fact,
one of the great advantages of the EOB approach is that, being Hamiltonian-based, it
provides a full set of evolution equations for all the dynamical variables:
$\mbox{\boldmath$x$}$, $\mbox{\boldmath$p$}$, $\mbox{\boldmath$S$}_1$ and
$\mbox{\boldmath$S$}_2$. See Eqs.~(3.1)--(3.4) of \cite{D01}. Here, we are interested in
comparing the EOB predictions with the numerical results of \cite{GGB1,GGB2} which are
restricted to the simple case where $\mbox{\boldmath$S$}_1$ and $\mbox{\boldmath$S$}_2$
are aligned with the orbital angular momentum $\mbox{\boldmath$L$}$. In this case, it
was shown in \cite{D01} that, for a given effective energy $E_{\rm eff}$, and a given
angular momentum $L$, the (effective) radius $r = \vert \mbox{\boldmath$x$} \vert$ of
the circular orbits is obtained by solving the two equations, $R (r,E_{\rm eff},L) =
\frac{\partial}{\partial r} \, R (r,E_{\rm eff},L) =0$, where $R$ is a certain effective
radial potential. [We use the fact that the Carter-like constant ${\cal Q}$ vanishes for
the ``equatorial'' orbits that we are considering.] Actually, it is more convenient to
work with the inverse radial variable $u \equiv M/r$ (we set $G=1$) and with the
corresponding $u$-potential $U(u) = \mu^{-2} \, r^{-4} \, R(r)$. Let us also introduce
the following dimensionless variables (note that $L \equiv L_z$ for the orbits we
consider and that, by definition, $L = L_{\rm eff} = L_{\rm real}$)
\begin{equation}
\label{eq2.4}
\widehat{E} \equiv \frac{E_{\rm real}}{\mu} \ , \quad \widehat{L} \equiv \frac{L}{\mu M}
\ , \quad \widehat{a} \equiv \frac{a}{M} \, ,
\end{equation}
and
\begin{equation}
\label{eq2.5}
\overline{E} \equiv \frac{E_{\rm eff}}{\mu} \ , \quad \overline{L} \equiv \frac{L - a \,
E_{\rm eff}}{\mu M} \equiv \widehat{L} - \widehat{a} \, \overline{E} \, .
\end{equation}
The quantity $a$ entering these equations is the modulus of the effective Kerr parameter
defined by Eq.~(\ref{eq2.3}).

In terms of these quantities the (scaled) $u$-potential reads
\begin{equation}
\label{eq2.6}
U(u) = (\overline{E} - \widehat{a} \, \overline{L} \, u^2)^2 - \overline{A} (u) \, (1 +
\overline{L}^2 \, u^2) \, ,
\end{equation}
where the function $\overline{A} (u) \equiv A(u) + \widehat{a}^2 \, u^2$ will be
discussed below.

The sequence of circular orbits is defined by the solutions of the two equations
\begin{equation}
\label{eq2.7}
U (u , \overline{E}, \overline{L}) = 0 = \frac{\partial}{\partial u} \, U (u,
\overline{E}, \overline{L}) \, .
\end{equation}
The effective Hamiltonian $\overline{E} = \overline{H}_{\rm eff} (u, \widehat{L})$ would
be obtained by solving $U (u , \overline{E}, \widehat{L} - \widehat{a} \, \overline{E})
= 0$ which is a quadratic equation in $\overline{E}$. However, the intrinsic simplicity
of the EOB approach is more visible if we use the variable $\overline{L} \equiv
\widehat{L} - \widehat{a} \, \overline{E}$ instead of $\widehat{L} \equiv L / \mu M$.
Indeed, the solution of $U (u , \overline{E}, \overline{L}) = 0$ reads
\begin{equation}
\label{eq2.8}
\overline{E} = \widehat{a} \, \overline{L} \, u^2 + \sqrt{\overline{A} (u) (1 +
\overline{L}^2 \, u^2)} \, .
\end{equation}
This result can be viewed as a simple modification by spin-orbit $(\widehat{a} \,
\overline{L} \, u^2$ term) and spin-spin effects $(\widehat{a}^2 \, u^2$ contribution to
$\overline{A} (u) = A(u) + \widehat{a}^2 \, u^2$) of the radial potential for circular
orbits in a spherically symmetric gravitational potential (such as a Schwarzschild
metric, or the effective metric in absence of spins): $\overline{E}_0 (u,\widehat{L}) =
\sqrt{A(u) (1 + \widehat{L}^2 \, u^2)}$. In such a situation $A(u)$ would simply be
(minus) the time-time component of the (effective) metric ($1-2u$ in the Schwarzschild
case). A nice feature of the EOB approach is that, in the spin-independent case, the
dynamics of circular orbits is fully encoded in only one function: the function $A(u) =
- g_{00}^{\rm eff} (u)$, with $u = M/r$. [As discussed in \cite{BD00,DJS00,D01}, the
dynamics of non circular orbits involves two more functions: $D(u)$ and $Q_4 (u,p)$.] It
is truly remarkable that the hundreds of complicated contributions entering the
PN-expanded two-body equations of motion get drastically compactified in so few
functions. In particular, for circular orbits, the essential physical information of the
PN-expanded dynamics is contained in the effective metric coefficient
\begin{equation}
\label{eq2.9}
A(u) = 1-2u + 2\nu \, u^3 + a_4 (\nu) \, u^4 + {\cal O} (u^5) \, .
\end{equation}
Note that there are no 1PN ($\propto u^2$) contribution in $A(u)$, and that the 2PN
effects amount to the very simple term $+ \, 2 \nu \, u^3$ \cite{BD00}. The 3PN
contribution $\propto u^4$ was found \cite{DJS00} to have the coefficient
\begin{equation}
\label{eq2.10}
a_4 (\nu) = \left( \frac{94}{3} - \frac{41}{32} \, \pi^2 + 2 \, \omega_s \right) \, \nu
\, .
\end{equation}
Again, remarkable simplifications happened in the calculation of the 3PN effective
metric. The 3PN coefficient $a_4 (\nu)$ could, a priori, have involved contributions
proportional to $\nu^2$ and $\nu^3$. All such contributions canceled to leave a simple
final result $a_4 (\nu) \propto \nu$. The parameter $\omega_s$ entering
Eq.~(\ref{eq2.10}) (which had been left ambiguous by the first 3PN calculations) has
been recently determined to be simply $\omega_s = 0$ \cite{DJS01}. This yields a
numerical value for the general relativistic prediction for the 3PN coefficient equal
to:
\begin{equation}
\label{eq2.11}
a_4^{\rm GR} (\nu) \simeq 18.688 \, \nu \simeq 4.672 \, (4\nu) \, .
\end{equation}

\subsection{Representation of the basic EOB function $A(u)$}

The EOB method a priori determines (from the original, fully
PN-expanded dynamics) only the PN-expansion of the basic function $A(u)$, or of the
combination
\begin{equation}
\label{eq2.12}
\overline{A} (u) \equiv A(u) + \widehat{a}^2 \, u^2 = 1 - 2 \, u + \widehat{a}^2 \, u^2
+ 2 \nu \, u^3 + a_4 (\nu) \, u^4 + {\cal O} (u^5) \, ,
\end{equation}
entering the effective radial potential (\ref{eq2.6}).

In this respect let us emphasize the various senses in which the EOB approach is a
``resummation'' of the original, fully PN-expanded equations of motion. A first, crucial
``resummation'' feature of the EOB approach is the mapping of the original dynamical
variables $\mbox{\boldmath$q$}_1$, $\mbox{\boldmath$p$}_1$, $\mbox{\boldmath$q$}_2$,
$\mbox{\boldmath$p$}_2$ onto the effective variables $\mbox{\boldmath$x$}$,
$\mbox{\boldmath$p$}$. This mapping \cite{BD00,DJS00} is given as a complicated PN
expansion, and provides a first level of simplification and compactification of the
dynamics. A second feature consists in the fact that the real Hamiltonian is given by an
iterated square-root. For instance, in the spin-independent case
($\widehat{\mbox{\boldmath$p$}} \equiv \mbox{\boldmath$p$} / \mu$)
\begin{equation}
\label{eq2.13}
H_{\rm real} (\mbox{\boldmath$x$} , \widehat{\mbox{\boldmath$p$}}) = M \sqrt{1 + 2\nu
\left(-1 + \sqrt{A(r) \left( 1 + \widehat{\mbox{\boldmath$p$}}^2 + \left(
\frac{A(r)}{D(r)} - 1 \right) (\mbox{\boldmath$n$} \cdot
\widehat{\mbox{\boldmath$p$}})^2 + \widehat{Q}_4 (\widehat{\mbox{\boldmath$p$}})
\right)}\right)} \, .
\end{equation}
This iterated square-root structure is, by itself, a partial resummation because the
fully PN-expanded Hamiltonian, which has the form $H_{\rm real} (\widehat{p}) =$ $M
[ 1 + \nu \left( \frac{1}{2} \, \widehat{\mbox{\boldmath$p$}}^2 + \frac{M}{r}
\right) + \cdots +$ $c_8 \, \widehat{\mbox{\boldmath$p$}}^8 + \cdots + {\cal O}
(\widehat{\mbox{\boldmath$p$}}^{10})]$, would be obtained by re-expanding all
this square-root structure in powers of $\widehat{\mbox{\boldmath$p$}}^2$ and $M/r$.
Finally, on top of these two partial resummations, it is very natural, within the EOB
approach, to further resum $H_{\rm real}$ by replacing the PN-expanded results for
$A(u)$, Eq.~(\ref{eq2.9}), and/or $\overline{A} (u)$, Eq.~(\ref{eq2.12}), by some better
behaved expressions. Indeed, as we recalled above, in the spin-independent case $A(u)$
plays the basic role of replacing $- \, g_{00}^{\rm Schwarz} = 1 - 2 M/r$ in the radial
potential determining the circular orbits.

It is then natural to require that the ``exact'' effective $A(u)$ should have the same
qualitative behavior as $1-2u$, i.e. a simple zero at a finite value of $u$. More
generally, in the spin-dependent case $\overline{A} (u)$ is a $\nu$-deformation of the
Kerr function $r^{-2} \, \Delta_t (r) = r^{-2} (r^2 - 2 \, M r + a^2)$ which
determines the location of the horizon. It is therefore natural to require that the
effective $\overline{A} (u)$ have, like $1 - 2u + \widehat{a}^2 \, u^2$, a simple zero
at a finite value of $u$. To ensure such a qualitative behavior it is natural to
replace the straightforward Taylor expansion giving $\overline{A} (u)$ by a suitable
Pad\'e approximant (chosen so as to robustly imply the presence of a simple zero),
namely \cite{D01}
\begin{equation}
\label{eq2.14}
\overline{A} (u , \widehat{a}^2) = P_3^1 \, [1 - 2u + \widehat{a}^2 \, u^2 + 2 \nu \,
u^3 + a_4 (\nu) \, u^4] \, ,
\end{equation}
where $P_m^n$ denotes a Pad\'e of the $N_n / D_m$ type (where the indices denote the
degrees of the numerator and denominator).

\begin{figure}
\includegraphics[height=6cm]{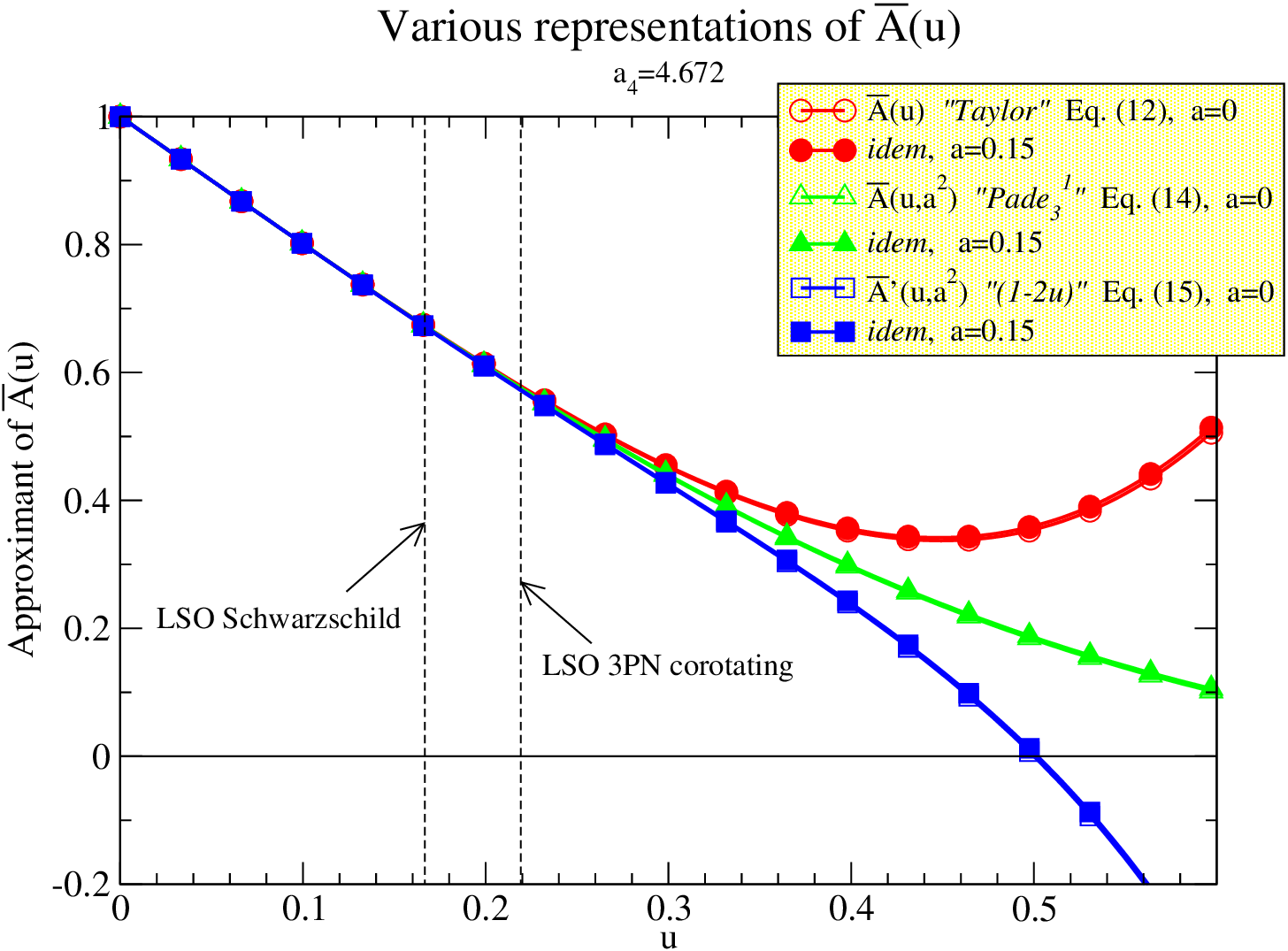}   \quad
\includegraphics[height=6cm]{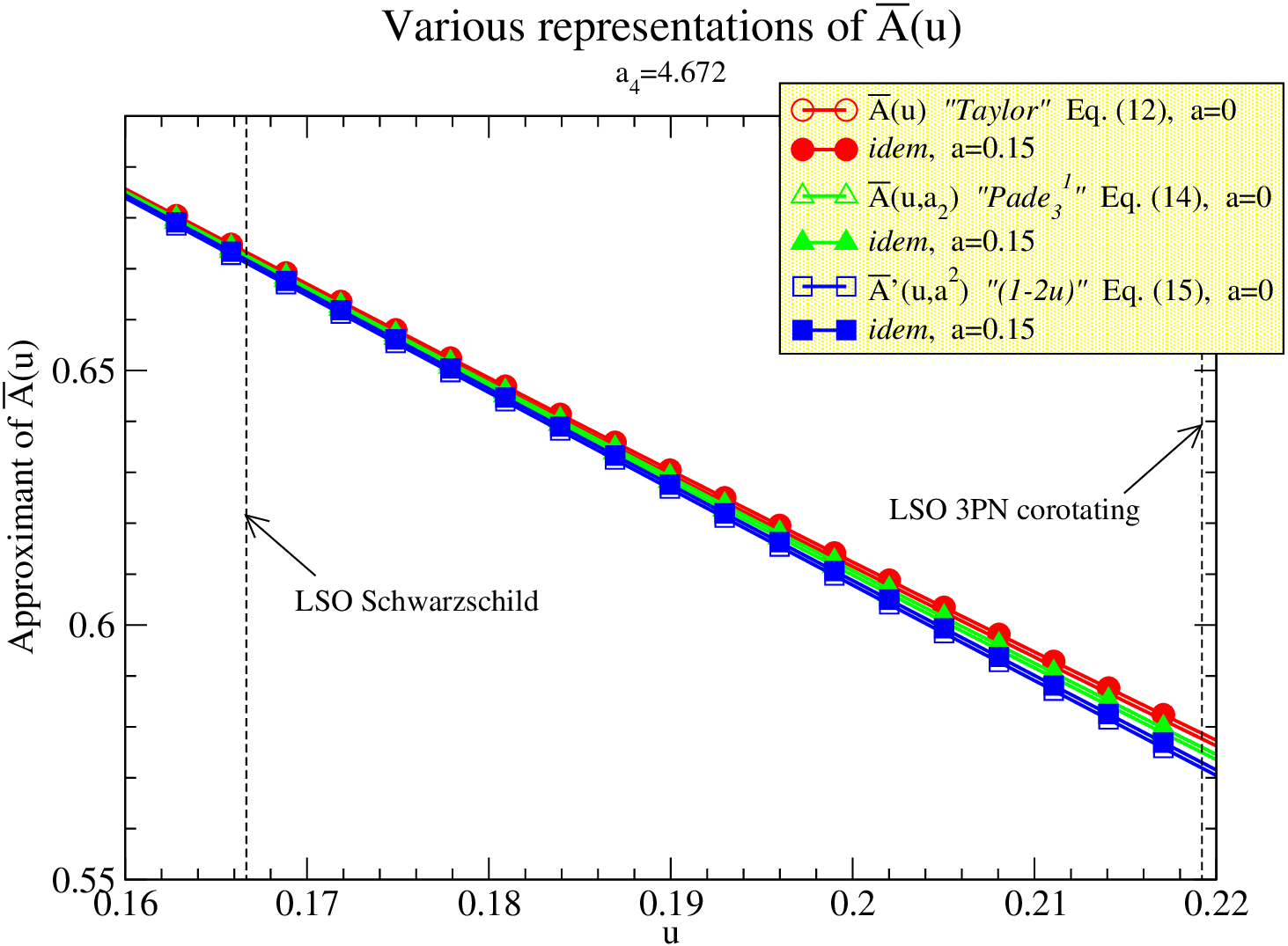}
\caption{\label{f:A(u)-var} Various representations of the function
$\overline{A} (u , \widehat{a}^2)$.
The right figure is a enlargement of the LSO region.}
\end{figure}

As a test of the robustness of the EOB predictions we shall also consider a more brutal
way of ensuring that $A(u)$ has a simple zero at a finite value of $u$: it consists in
defining
\begin{equation}
\label{eq2.15}
\overline{A}' (u , \widehat{a}^2) = (1-2u) (1 + 2 \nu \, u^3 + (a_4 (\nu) + 4 \nu) \,
u^4) + \widehat{a}^2 \, u^2 \, .
\end{equation}
Fig.~\ref{f:A(u)-var} compares the various representations of the function $\overline{A} (u)$: the
straightforward Taylor series approximant (\ref{eq2.12}) (truncated after the 3PN term
$\propto u^4$), the Pad\'e approximant (\ref{eq2.14}) and the ``factorized Taylor''
approximant (\ref{eq2.15}). They are represented both in the limit $\widehat{a}=0$,
and for the value $\widehat{a}=0.15$ which roughly corresponds to the value of
 $\widehat a$, at the LSO, in
the effective one-body formalism discussed in the next Section. Note that the various
representations start differing significantly from each other only for $u \simeq 0.4$,
i.e. for values about twice larger than the value of $u$ corresponding to the LSO, i.e.
$u_{\rm LSO} \simeq 0.22$. However, this does not mean that the resuming of
$\overline{A} (u)$ and/or the addition of the spin-dependent interactions has no
significant effect on the characteristics of the LSO. Indeed, as the usual (non
corotating) LSO corresponds to an inflection point in the effective radial potential
$H_{\rm eff} (r,L,S_a)$, any modification of the radial dependence of the Hamiltonian
(e.g. by a change in $\overline{A} (u)$ or the addition of the linear spin-orbit term
$\widehat a \, \overline L \, u^2$ in Eq.~(\ref{eq2.8})) has a strong effect on the
location of the LSO.

\begin{figure}
\includegraphics[height=6cm]{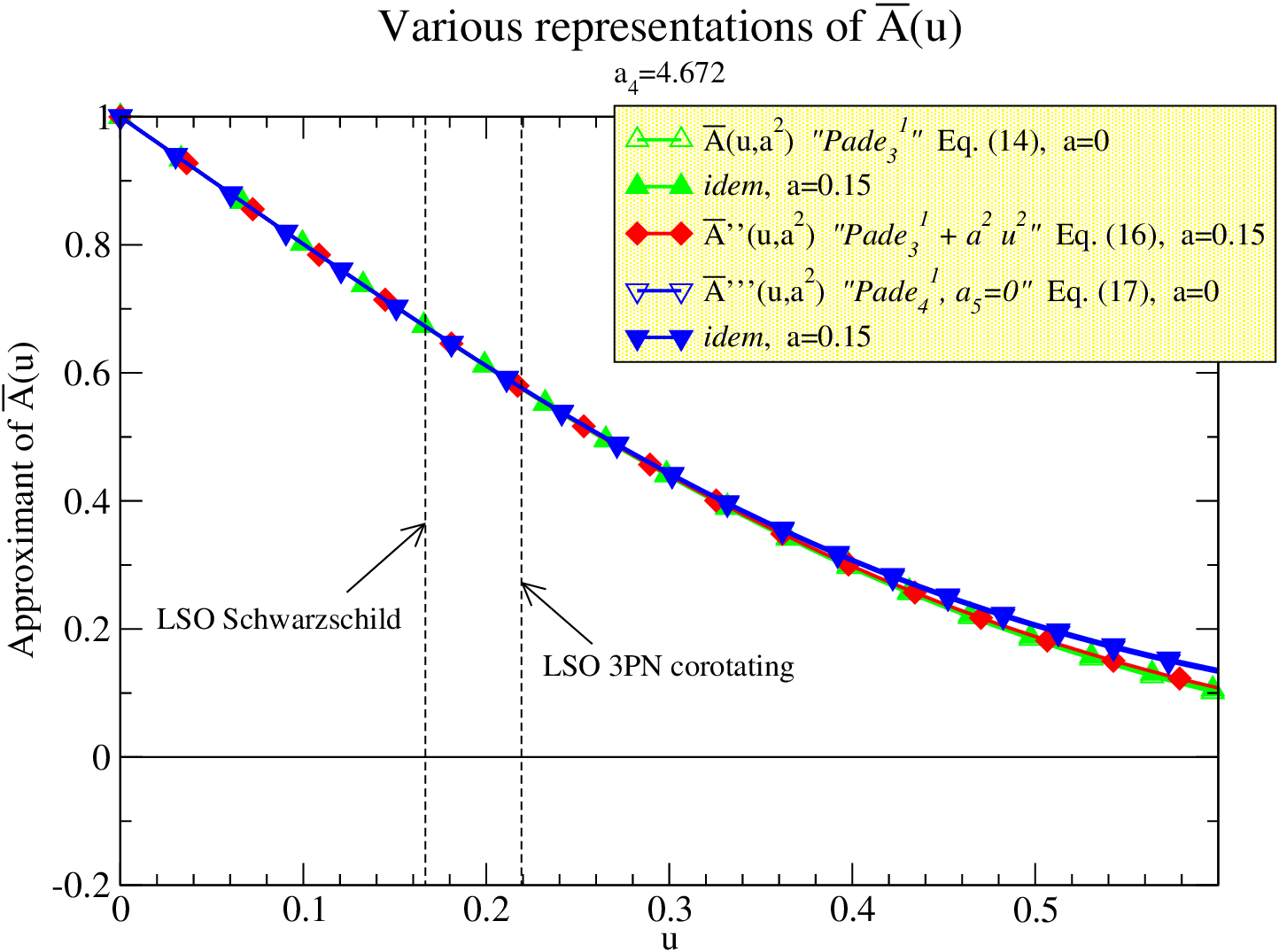}   \quad
\includegraphics[height=6cm]{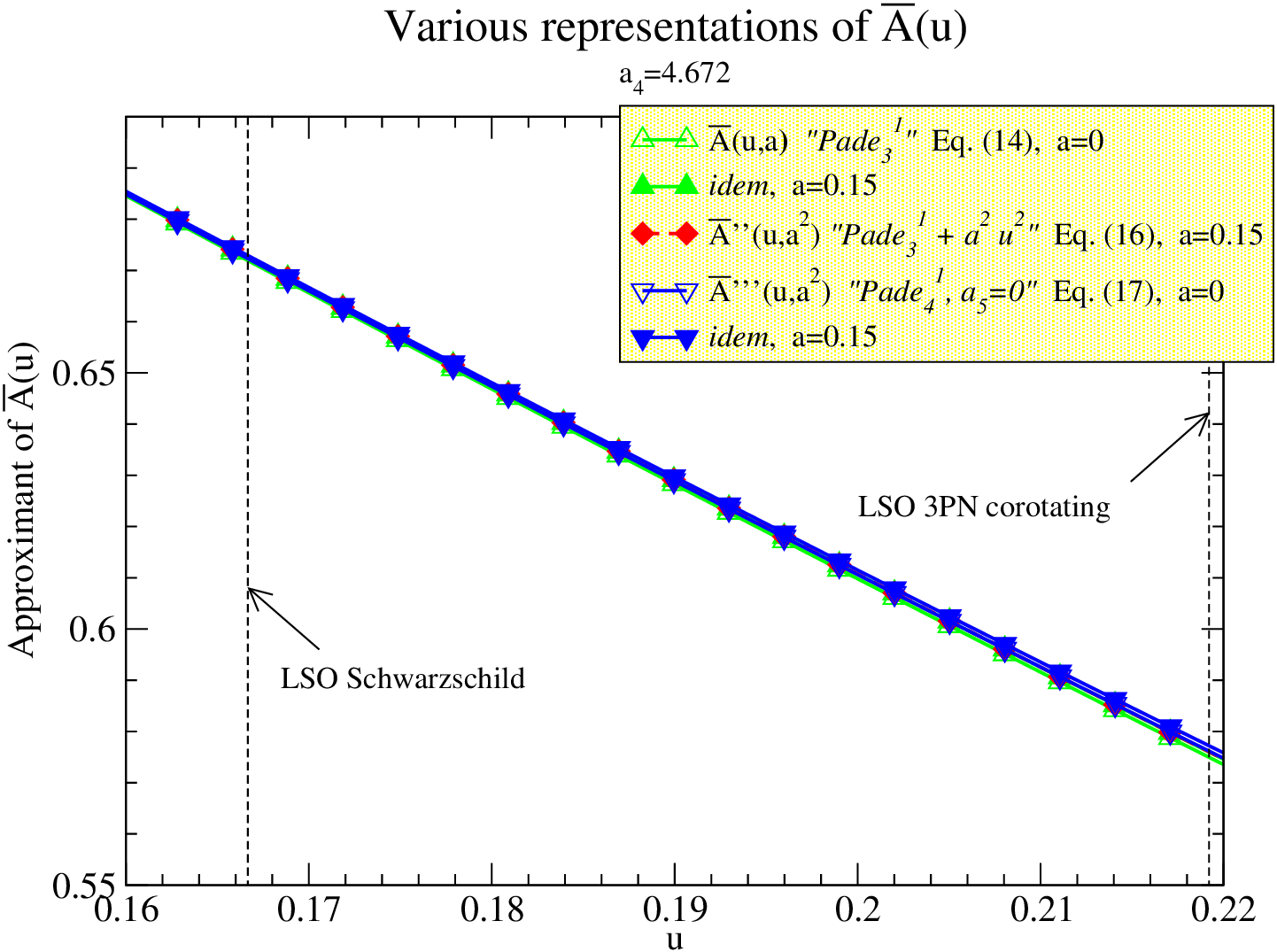}
\caption{\label{f:A(u)-Pade} Various Pad\'e approximants of the function
$\overline{A} (u , \widehat{a}^2)$.
The right figure is a enlargement of the LSO region.}
\end{figure}

On the other hand, Fig.~\ref{f:A(u)-Pade} compares different variants of the
Pad\'e approximants that
one could use. In the text, we have followed Ref.~\cite{D01} in focusing on the
specific definition (\ref{eq2.14}) of the function $\overline{A} (u)$. However, one
could have considered
\begin{equation}
\label{eq2.16}
\overline{A}'' (u , \widehat{a}^2) = P_3^1 [1 - 2u + 2\nu \, u^3 + a_4 (\nu) \, u^4] +
\widehat{a}^2 \, u^2 \, .
\end{equation}
Another  possibility arises when one uses the great ``flexibility'' of the EOB approach.
As discussed in \cite{BD00,D01} one can think of the EOB formalism as a multi-parameter
analytical formalism, where only a fraction of the parameters is currently known. For
instance, there are further terms in the expansions (\ref{eq2.9}) and (\ref{eq2.12}),
say $+ \, a_5 (\nu) \, u^5 + a_6 (\nu) \, u^6 + \cdots$. The parameters $a_5 (\nu) , a_6
(\nu) , \ldots$ are not known at present.
However, they can be introduced as {\it fitting parameters} and adjusted so as to
reproduce other information one has about the exact results. For instance, if we had
reason to trust some numerical results, one could search for the optimal values of $a_4
(\nu) , a_5 (\nu) , \ldots$ that best fit the numerical data. We shall give an example
of this below. A minimal requirement is to test the ``robustness'' of the EOB formalism
under adding some reasonable values for $a_5 (\nu) , \ldots$. See Ref.~\cite{DIJS} for a
study of this robustness. Here, we just wish to illustrate a minimal aspect of this
robustness: when adding the next term $a_5 (\nu) \, u^5$ to $A(u)$, the natural Pad\'e
approximant to use becomes
\begin{equation}
\label{eq2.17}
\overline{A}''' (u , \widehat{a}^2) = P_4^1 [1-2u + \widehat{a}^2 \, u^2 + 2 \nu \, u^3
+ a_4 (\nu) \, u^4 + a_5 (\nu) \, u^5 ] \, .
\end{equation}

If we take the $a_5 \rightarrow 0$ limit of the Pad\'e approximant (\ref{eq2.17}) we do
not recover (\ref{eq2.14}). Indeed, the $P_3^1$ approximant (\ref{eq2.14}) is uniquely
defined by the requirement of being $\propto N_1 / D_3$ and of matching (\ref{eq2.12})
up to $a_4 \, u^4$. This means that $P_3^1$ effectively makes an ``educated guess''
about an infinite number of additional contributions $a_5 \, u^5 + a_6 \, u^6 + \cdots$,
which are plausible continuations of the known terms $1 - 2u + \cdots + a_4 \, u^4$. In
particular, the Taylor expansion of $P_3^1$ must contain a non-zero contribution $a_5 \,
u^5$. This explains why it does not agree with the $a_5 = 0$ limit of (\ref{eq2.17}). It
is therefore interesting to compare (\ref{eq2.17}) to (\ref{eq2.14}) to measure the
robustness of Pad\'e approximants against knowledge or lack of knowledge of $a_5$. In
fact, Fig.~\ref{f:A(u)-Pade} shows that the various ways we have just described of defining some Pad\'e
approximants of $\overline A (u)$ give extremely close results. Much closer than the
variants represented in Fig.~\ref{f:A(u)-var}.
This is an illustration of the generic robustness of Pad\'e resummation.

\section{Comparing analytical and numerical results for corotating binary black
holes}\label{sec3}

\subsection{Corotating configurations in the EOB framework}

Let us now tackle the central problems of this paper: (i) to compute the
coordinate-invariant functions characterizing the dynamics of adiabatic sequences of
corotating binary black holes on circular orbits, and (ii) to compare the (EOB)
analytical predictions with the numerical results of Refs.~\cite{GGB1,GGB2}. The main
subtlety in this study comes from the {\it corotation} assumption. This assumption means
that the spins of the black holes vary along the sequence, and in fact increase as the
radial distance $r$ decreases. However, any point in the sequence, i.e. any circular
orbit, must be determined (in the EOB approach) by imposing that
\begin{equation}
\label{eq3.1}
0 = \dot r = \frac{\partial H_{\rm real} (r,p_r , L,S_1 , S_2)}{\partial r} \, ,
\end{equation}
\begin{equation}
\label{eq3.2}
0 = \dot{p}_r = - \frac{\partial H_{\rm real} (r,p_r , L,S_1 , S_2)}{\partial p_r} \, ,
\end{equation}
and $p_r = 0$, where the derivatives are all considered for {\it fixed} values of $L$,
$S_1$ and $S_2$. Eq.~(\ref{eq3.2}) is automatically satisfied by $p_r = 0$, and
Eq.~(\ref{eq3.1}) (with $p_r = 0$) then leads to Eqs.~(\ref{eq2.7}) above.
Eq.~(\ref{eq3.1}) determines $r$ as a function of $L$, $S_1$ and $S_2$, or $L$ as a
function of $r$, $S_1$ and $S_2$. It then remains to determine how $S_1$ and $S_2$ vary
as functions of $L$ and $r$. Let us also note beforehand that, in view of
Eq.~(\ref{eq3.1}), the usual (irrotational) LSO corresponds to an inflection point
 (in $r$) of $H_{\rm real} (r,L,S_a)$ considered for fixed values of $L$ and $S_a$
(corresponding to the LSO), and to a minimum of $H_{\rm real}$ (i.e. a maximum binding
configuration) along a sequence of fixed values of $S_a$. In the corotating case, we
define the LSO as the maximum binding configuration along a corotating sequence.
The variation of $S_a$ with
$r$ along a corotating sequence changes the location of the minimum of
$H_{\rm real}$ with respect to the irrotational case. The variation of $S_a$ with
$r$ prevents this minimum to correspond to an inflection point of the function
$H_{\rm real} (r,L,S_a)$ considered for the fixed LSO values of $L$ and $S_a$.

The variation of the magnitudes of the spin vectors along a sequence obliges us to
complete the formalism of \cite{D01} by giving a prescription for the variation of the
black hole masses $m_1$ or $m_2$ with $\mbox{\boldmath$S$}_1^2$ or
$\mbox{\boldmath$S$}_2^2$. [This issue did not show up in \cite{D01} because the EOB
formalism guaranteed the conservation of $\mbox{\boldmath$S$}_1^2$ and
$\mbox{\boldmath$S$}_2^2$.] Though, in principle, there are subtleties linked to the
exact physical meaning of the EOB spin vectors, at the level of approximation of
Ref.~\cite{D01} it is clear that we can simply require that each gravitational (and
inertial) mass $m_a$ (with $a=1,2$) depends on $\mbox{\boldmath$S$}_a^2$ essentially as
in the Christodoulou mass formula \cite{X}
\begin{equation}
\label{eq3.3}
m_a [m_a^{\rm irr} , \mbox{\boldmath$S$}_a] = \sqrt{(m_a^{\rm irr})^2 +
\frac{\mbox{\boldmath$S$}_a^2}{4 (m_a^{\rm irr})^2}} \, .
\end{equation}
Here, the irreducible mass $m_a^{\rm irr}$ is an adiabatic invariant (constant along a
corotating sequence), which should be related to the area of the horizon via $A_a = 16
\pi (m_a^{\rm irr})^2$.

The simplest definition of a corotating sequence is that the spin angular velocities of
the black holes $\Omega_a$ stay equal to the orbital angular velocity $\Omega_0$:
\begin{equation}
\label{eq3.4}
\Omega_1 = \Omega_2 = \Omega_0 \, .
\end{equation}

The universal relations of (Hamiltonian) dynamics (such as $\Omega_0 = d \varphi / dt =
+ \, \partial H_{\rm real} / \partial p_{\varphi}$ with $p_{\varphi} = L_z = L$) tell us
that the angular velocities are obtained (for circular orbits with aligned spins) as
\begin{equation}
\label{eq3.5}
\Omega_0 = \frac{\partial H_{\rm real} (r,L,S_1 , S_2)}{\partial L} \, , \ \Omega_a =
\frac{\partial H_{\rm real} (r,L,S_1 , S_2)}{\partial S_a} \, ,
\end{equation}
where $H_{\rm real} (r,L,S_a)$ is the full (real) Hamiltonian,
\begin{equation}
\label{eq3.6}
H_{\rm real} = H_{\rm real} (m_a^{\rm irr} , r , p_r , p_{\varphi} , S_a) \, ,
\end{equation}
considered for fixed values of the irreducible masses $m_a^{\rm irr}$, and for $p_r = 0$
(and $p_{\varphi} = L_z = L$). Note that this means, in particular, that in expressions
such as Eq.~(\ref{eq2.2}) the masses $m_1 , m_2$, as well as $M = m_1 + m_2$ and $\nu =
m_1 \, m_2 / (m_1 + m_2)^2$, must be considered as functions of $m_a^{\rm irr}$ and
$S_a$. The dependence of the masses on $S_a$ is crucial in the computation of the spin
angular velocities $\Omega_a = \partial H_{\rm real} / \partial S_a$. In fact, one can
see that $\Omega_a$ is the sum of two contributions: a ``bare'' contribution
$\Omega_a^{\rm bare} = \partial m_a / \partial S_a$ which is the angular velocity of an
isolated Kerr hole, and a contribution linked to the presence of the other hole, which
represents a frame-dragging effect, which is automatically included in the EOB approach.
[Note the consistency of the EOB approach under the replacement (\ref{eq3.3}): this
yields a new (numerically dominant) contribution to
${\bf \Omega}_a = \frac{\partial H_{\rm real} }{\partial {\bf S}_a}$ which drops out
of the spin-evolution equation (3.5) of Ref. \cite{D01}. Note also that one would need
to introduce extra (dissipative) terms in the spin evolution equation to enforce
a realization of a corotating sequence in the EOB formalism.]

Along any continuous sequence of circular orbits (not necessarily a corotating one) one
has $\partial H_{\rm real} / \partial r = 0 = \partial H_{\rm real} / \partial p_r$ (see
Eqs. (\ref{eq3.1}), (\ref{eq3.2})), so that we can write the variation of the energy as
\begin{equation}
\label{eq3.7}
d E_{\rm real} = \Omega_0 \, d L + \Omega_1 \, d S_1 + \Omega_2 \, d S_2 \, .
\end{equation}

Note that, in the special case of a corotating sequence we have
\begin{equation}
\label{eq3.8}
d E_{\rm real} = \Omega \, d J
\end{equation}
where $J = L + S_1 + S_2$ is the conserved total angular momentum of the EOB approach
(see Eq.~(3.7) of \cite{D01}), and where $\Omega$ is the common value of the angular
velocities (\ref{eq3.4}).

Let us, for a moment, shift from the analytical side to the numerical side. One of the
nice advantages of the numerical approach of \cite{GGB1,GGB2} is that it determines not
only the global (Arnowitt-Deser-Misner; ADM) conserved quantities $M_{\rm ADM}$ and
$J_{\rm ADM}$, but also the common angular velocity $\Omega_K$ (which appears in the
asymptotic boundary condition for the helical Killing vector $\ell^{\mu}
\partial_{\mu}$). A corotating sequence is then numerically defined by imposing the
satisfaction of $d M_{\rm ADM} = \Omega_K \, d J_{\rm ADM}$. In view of Eq.~(\ref{eq3.8}),
it is then fully consistent to identify the numerical and analytical quantities
according to $M_{\rm ADM} = E_{\rm real}$, $\Omega_K = \Omega$, $J_{\rm ADM} = J$.
Ref.~\cite{GGB2} has also defined a total (numerical) irreducible mass $M_{\rm irr}$ as
the sum $\sqrt{A_1 / 16 \pi} + \sqrt{A_2 / 16 \pi}$, and has checked that it remains
constant along a corotating sequence. It is therefore fully consistent to identify
$M_{\rm irr}$ with the analytical quantity $m_1^{\rm irr} + m_2^{\rm irr}$. [Once the
constancy of $M_{\rm irr}$ is numerically verified, the consideration of $M_{\rm irr}$
for infinitely separated configurations, for which the areas $A_a$ unambiguously tend to
their Kerr values in both formalisms, forces the identification $M_{\rm irr} = m_1^{\rm
irr} + m_2^{\rm irr}$.]

The numerical results of \cite{GGB2} give access to several invariant functions, such as
$M_{\rm ADM} / M_{\rm irr}$ as a function of $\Omega_K M_{\rm irr}$, and $J_{\rm ADM} /
M_{\rm irr}^2$ as a function of $\Omega_K M_{\rm irr}$. However, let us mention that this
normalization with respect to $M_{\rm irr}$ is different from the one used in \cite{GGB2},
where the global quantities are chosen so that $M_{\rm ADM}=1$ at the LSO (normalization
with $M_0$ in the language of \cite{GGB2}).
 Now that we have shown that
these quantities have unambiguous correspondents in the (EOB) analytical framework, it
remains to complete the analytical determination of the corotating sequence.

To do this we need to write explicitly the equations (\ref{eq2.7}) and (\ref{eq3.4}).
Let us first consider (\ref{eq2.7}), i.e. the conditions defining circular orbits. It is
convenient to divide the radial potential $U(u)$ by $\overline A (u)$ before
differentiating with respect to $u$. Let us introduce the short-hand notation
\begin{eqnarray}
\label{eq3.8bis}
\alpha (u , \widehat a ) &\equiv &\frac{1}{\overline A (u , \widehat a )} \, , \nonumber
\\
\beta (u , \widehat a ) &\equiv &\frac{\widehat a \, u^2}{\overline A (u , \widehat a )}
\, , \nonumber \\
\gamma (u , \widehat a ) &\equiv &u^2 - \frac{\widehat{a}^2 \, u^4}{\overline A (u ,
\widehat a )} \, .
\end{eqnarray}

Eqs.~(\ref{eq2.7}) are then equivalent to the two equations
\begin{subequations}
\label{eq3.9}
\begin{eqnarray}
\alpha \, \overline{E}^2 - 2 \, \beta \, \overline E \, \overline L &= &\gamma \,
\overline{L}^2 + 1 \, ,
\label{eq3.9a} \\
\alpha' \, \overline{E}^2 - 2 \, \beta' \, \overline E \, \overline L &= &\gamma' \,
\overline{L}^2 \, ,
\label{eq3.9b}
\end{eqnarray}
\end{subequations}
where the prime denote a derivative with respect to $u$. Eq.~(\ref{eq3.9b}) is
homogeneous (of second degree) in $\overline E$ and $\overline L$. We can therefore get
a second degree equation for the ratio $\lambda \equiv \overline L / \overline E$. With
sign conventions such that $L = L_z > 0$, one finds that $\lambda$ is determined as
\begin{equation}
\label{eq3.10}
\lambda = \lambda_2 (u,\widehat a) \equiv \frac{-\beta' + \sqrt{{\beta'}^2 + \alpha'
\gamma'}}{\gamma'} \, .
\end{equation}
Inserting this solution in Eq.~(\ref{eq3.9a}) then determines $\overline{E}^2$. Finally,
we can write $\overline E$ and $\overline L$ as the following explicit functions of $u$
and $\widehat a$:
\begin{subequations}
\label{eq3.11}
\begin{eqnarray}
\overline E = \overline{E}_2 (u,\widehat a) &\equiv &(\alpha - 2 \, \beta \, \lambda -
\gamma \, \lambda^2)^{-1/2} \, ,
\label{eq3.11a} \\
\overline L = \overline{L}_2 (u,\widehat a) &\equiv &\lambda (\alpha - 2 \, \beta \,
\lambda - \gamma \, \lambda^2)^{-1/2} \, ,
\label{eq3.11b}
\end{eqnarray}
\end{subequations}
where $\lambda$ must be replaced by Eq.~(\ref{eq3.10}). In the equations above we have
added the index 2 to various functions to indicate that they are functions of the two
variables $u$ and $\widehat a$. At this stage, the results are valid for any circular
orbit, not necessarily a member of a (specific) corotating sequence. Below, we shall
determine how the spins, and thereby $\widehat a$, vary along a corotating sequence,
i.e. vary as functions of $u$. This will give rise to functions of only one variable:
$u$.

{}From the results (\ref{eq3.11}) one can determine the (scaled) real orbital angular
momentum $\widehat L$, and the (scaled) real energy $\widehat E$ [see
Eqs.~(\ref{eq2.2}), (\ref{eq2.4}) and (\ref{eq2.5})]. Explicitly
\begin{subequations}
\label{eq3.12}
\begin{eqnarray}
\widehat E = \widehat{E}_2 (u,\widehat a) &\equiv &\frac{1}{\nu} \sqrt{1 + 2 \nu
(\overline{E}_2 (u,\widehat a) - 1)} \, , \label{eq3.12a} \\
\widehat L = \widehat{L}_2 (u,\widehat a) &\equiv &\overline{L}_2 (u,\widehat a) +
\widehat a \, \overline{E}_2 (u,\widehat a) \, . \label{eq3.12b}
\end{eqnarray}
\end{subequations}
To write explicitly the condition (\ref{eq3.4}) let us introduce the scaled orbital
angular velocity
\begin{equation}
\label{eq3.13}
\widehat{\Omega}_0 \equiv M \, \Omega_0 \, .
\end{equation}
When $\widehat a$, and the spins, are fixed, Eq.~(\ref{eq3.7}) gives $d \,
\widehat{E}_{\rm real} = \widehat{\Omega}_0 \, d \, \widehat L$ so that we can compute
$\widehat{\Omega}_0$ as follows from Eqs.~(\ref{eq3.12}):
\begin{equation}
\label{eq3.14}
\widehat{\Omega}_0 (u,\widehat a) = \frac{\partial \, \widehat{E}_2 (u,\widehat a) /
\partial u}{\partial \, \widehat{L}_2 (u,\widehat a) / \partial u} \, .
\end{equation}
Note that, at the LSO (corresponding to any fixed value of $\widehat a$), both the
denominator and the numerator of Eq.~(\ref{eq3.14}) have a simple zero, so that
$\widehat{\Omega}_0$ has a well defined limit.

The most complicated calculation is that of the spin angular velocities $\Omega_1$ and
$\Omega_2$. Indeed, the dependence of $E_{\rm real}$ on $S_a$ is quite involved as,
besides the explicit dependence on the spins, one must also consider the implicit spin
dependence (via Eq.~(\ref{eq3.3})) of all the mass-dependent quantities: $M$, $\mu$,
$\nu$, $\sigma_1$, $\sigma_2, \ldots$ The simplest way to do the calculation is to use
the general identity (\ref{eq3.7}). A simplification comes from the fact that, following
\cite{GGB2}, we can restrict ourselves to the {\it symmetric case}: $m_1 = m_2$, and $S_1
= S_2$. For such a case, it is easily seen that, when differentiating functions which
are $1 \leftrightarrow 2$ symmetric (such as $E_{\rm real}$) it is correct to consider
quantities such as $\nu = m_1 \, m_2 / M^2$, $\sigma_1 = 1 + 3 \, m_2 / (4 \, m_1)$,
$\sigma_2 = 1 + 3 \, m_1 / (4 \, m_2)$ as fixed when varying the spins. Setting
\begin{equation}
\label{eq3.15}
\widehat{\Omega}_1 \equiv M \, \Omega_1 \, , \ \widehat{a}_1 \equiv \frac{S_1}{m_1^2} =
\frac{8}{7} \, \widehat a \, ,
\end{equation}
a long, but straightforward, calculation leads to the following expression for
$\widehat{\Omega}_1$:
\begin{equation}
\label{eq3.16}
\widehat{\Omega}_1 = \frac{\widehat{a}_1}{1 + \sqrt{1 - \widehat{a}_1^2}} \left(
\frac{1}{4} \, \widehat{E}_2 - \frac{1}{2} \, \widehat{\Omega}_0 \, \widehat{L}_2
\right) + \frac{7}{16} \, \sqrt{1 - \widehat{a}_1^2} \left( \frac{\partial  \,
\widehat{E}_2}{\partial \, \widehat{a}} - \widehat{\Omega}_0 \, \frac{\partial \,
\widehat{L}_2}{\partial \, \widehat a} \right) \, .
\end{equation}
The functions $\widehat{E}_2$ and $\widehat{L}_2$ are that defined in
Eqs.~(\ref{eq3.12}) above. Let us recall that Eq.~(\ref{eq3.16}) is valid in the
symmetric case where $\widehat{a}_1 = \widehat{a}_2 = (8/7) \, \widehat{a}$,
$\sigma_1 = \sigma_2 = 7/4$, $\nu = 1/4$, $\widehat{\Omega}_1 = \widehat{\Omega}_2$,
etc$\ldots$

When we impose the corotating condition $\widehat{\Omega}_1 = \widehat{\Omega}_0$,
Eq.~(\ref{eq3.16}), yields an equation to determine
$\widehat{a}_1 \equiv (8/7) \, \widehat{a}$
as a function of $u$. For instance, one can rewrite Eq.~(\ref{eq3.16}) in the form
$\widehat{a} = f (u,\widehat{a})$. As $\widehat{a}$ is a rather small quantity, we can
solve for $\widehat{a}$ by iteration: e.g., with sufficient accuracy, $\widehat{a}
\simeq f (u,f(u,f(u,0)))$. Having so obtained $\widehat{a}$ as a function of $u$, say
$\widehat{a} = \widehat{a}_u (u)$, we can then insert this result in the previous
expressions to derive a ``basis'' of independent (dimensionless) quantities, say:
\begin{eqnarray}
\label{eq3.17}
\widehat E &= &\widehat{E}_1 (u) \equiv \widehat{E}_2 (u , \widehat{a}_u (u)) \, ,
\nonumber \\
\widehat L &= &\widehat{L}_1 (u) \equiv \widehat{L}_2 (u , \widehat{a}_u (u)) \, ,
\nonumber \\
\widehat{\Omega}_0 &= &\widehat{\Omega}_0 (u , \widehat{a}_u (u)) \, , \nonumber \\
\widehat{a}_1 &= &\frac{8}{7} \, \widehat{a}_u (u) \, , \nonumber \\
\frac{m_1}{m_1^{\rm irr}} &= &\sqrt{\frac{2}{1 + \sqrt{1 - \widehat{a}_1^2}}} \, .
\end{eqnarray}

{}From this ``basis'' we can then compute, as functions of $u$, all the quantities we
might need, and, in particular, the dimensionless quantities which are most directly
derived from the numerical calculations, namely
\begin{eqnarray}
\label{eq3.18}
\frac{M_{\rm ADM}}{M_{\rm irr}} &= &\frac{E_{\rm real}}{m_1^{\rm irr} + m_2^{\rm irr}} =
\frac{1}{4} \, \frac{m_1}{m_1^{\rm irr}} \, \widehat E \, , \nonumber \\
\frac{J_{\rm ADM}}{M_{\rm irr}^2} &= &\frac{L + S_1 + S_2}{(m_1^{\rm irr} + m_2^{\rm
irr})^2} = \frac{1}{4} \left( \frac{m_1}{m_1^{\rm irr}} \right)^2 \, [\widehat L + 2 \,
\widehat{a}_1] \, , \nonumber \\
M_{\rm irr} \, \Omega_K &= &\frac{m_1^{\rm irr}}{m_1} \, \widehat{\Omega}_0 \, .
\end{eqnarray}

\begin{figure}
\includegraphics[height=8cm]{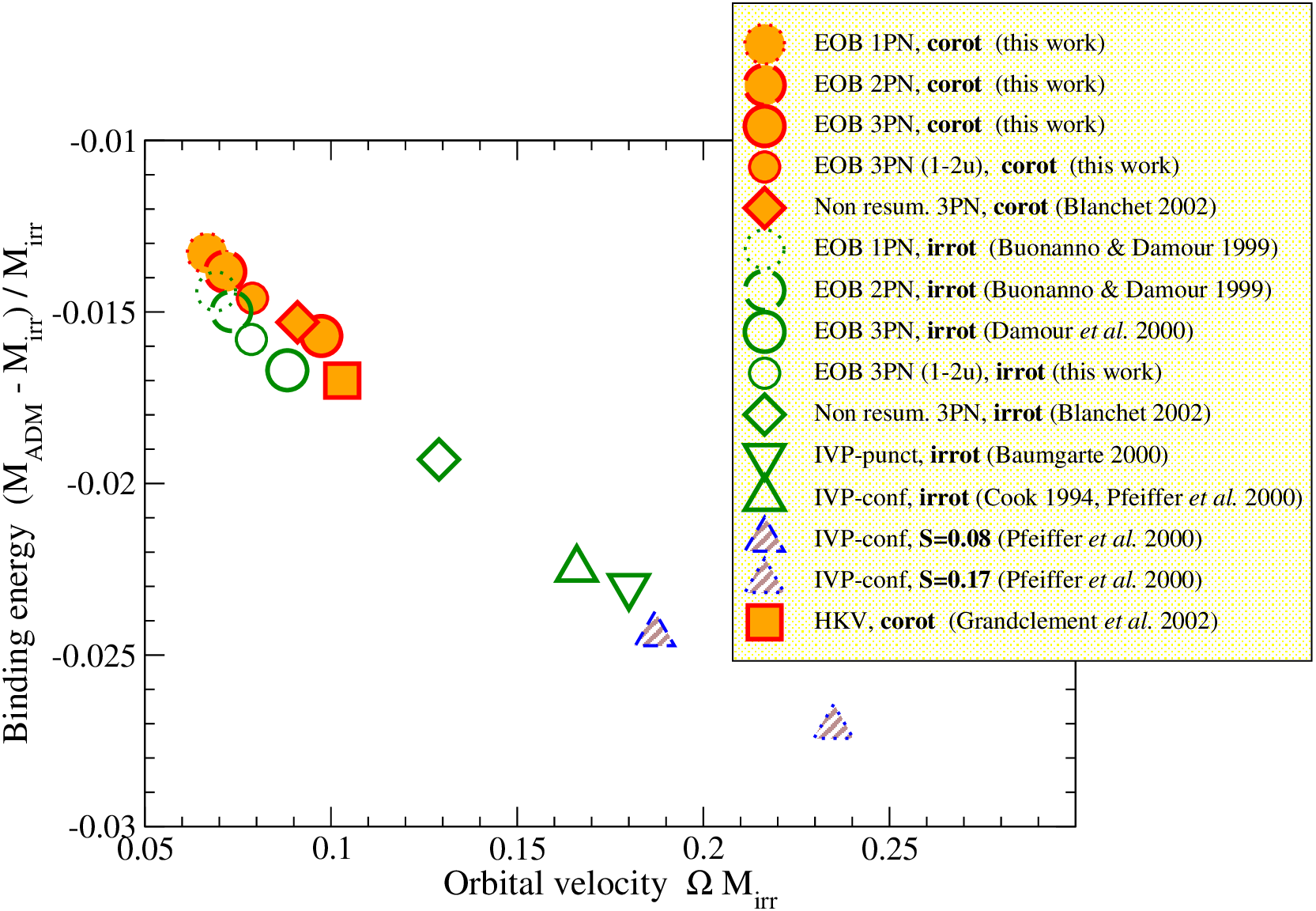}
\caption{\label{f:comp_lso} Maximum binding energy configurations
(LSO) obtained with various analytical and numerical methods.
Empty (resp. filled) symbols denote irrotational (resp. corotating) systems.
`EOB 3PN' denotes the computation performed with the Pad\'e approximant
$\overline{A}(u,{\hat a}^2)$ given by Eq.~(\ref{eq2.14}), whereas `EOB 3PN (1-2u)'
corresponds to the $\overline{A}'(u,{\hat a}^2)$ form given by Eq.~(\ref{eq2.15}).
References are as follows: Blanchet 2002 \cite{B02}; Buonanno \& Damour
1999 \cite{BD00}; Damour et al. 2000 \cite{DJS00}; Baumgarte 2000 \cite{baumgarte};
Cook 1994 \cite{cook}; Pfeiffer et al. 2000 \cite{PfeifTC00};
Grandcl\'ement et al. \cite{GGB2}.}
\end{figure}

\subsection{Characteristics of the LSO}

Figure~\ref{f:comp_lso} compares various analytical and numerical estimates of the
{\it maximum} binding energy
$(E_{\rm real} / M_{\rm irr}) - 1$, together with the corresponding dimensionless angular
velocity $M_{\rm irr} \, \Omega$, along sequences of circular orbits of equal-mass black
holes. Note that some sequences are corotating while others are irrotational. Several
interesting lessons can be drawn from Fig.~\ref{f:comp_lso}.
The first one is that the analytical (EOB) results
show that the effect of the corotation is quite small. [The reason for this smallness
is discussed in detail below.] In fact, the corotation effects are smaller or comparable
to the differences induced by making other changes in the treatment of the problem (like
using a different resummation method for the basic EOB potential $A(u)$, or working at the 2PN
approximation instead of the 3PN one).

Another important lesson drawn from Fig.~\ref{f:comp_lso}
concerns the PN robustness of the EOB estimates (i.e. their stability under a change of
PN accuracy): Indeed, we have
compared the results corresponding to the 1PN approximation of the EOB potential
$\overline{A} (u)$, namely
\begin{equation}
\overline{A}^{\rm 1PN} (u) = 1 - 2u + \widehat{a}^2 \, u^2 \, ,
\end{equation}
to its 2PN approximation
\begin{equation}
\overline{A}^{\rm 2PN} (u) = P_2^1 [1-2u + \widehat{a}^2 \, u^2 + 2 \, \nu \, u^3] \, ,
\end{equation}
and to its 3PN approximation (\ref{eq2.14}) (with $a_4(\nu)$ given by Eq.(\ref{eq2.11})).

We see that, on the scale of the differences between the various estimates, and in
particular, on the scale of the differences of the various numerical results among
themselves, the EOB method yields rather close results at all PN approximations.
This is one of the signs of the good resummation properties of the EOB method, and is to be
contrasted with the much poorer behavior (under a change of PN accuracy)
 of the other analytical methods of estimating
LSO characteristics such as: the direct use of PN-expanded equations of motion \cite{LW90}
(as discussed in Table~II of \cite{KWW93}), the use of PN-expanded Hamiltonians (as discussed
in \cite{WS} and \cite{Buona02}; see, e.g., Fig.~3 of \cite{Buona02}),
the $E$-method (discussed in Table~I, Table~VII
and Fig.~5 of \cite{DIS98}, and in Fig.~3 of \cite{B02}), the $e$-method (discussed in
\cite{DIS98} and \cite{DJS00}) or the $j$-method \cite{DJS00}.

For comparison purposes, we have included in Fig.~\ref{f:comp_lso} the (irrotational and
corotating) analytical predictions recently derived by using the the non-resummed
$E$-method at the 3PN level \cite{B02}. We note that the prediction of the non-resummed
$E$-method for the {\it irrotational} LSO angular velocity is $46$\% {\it larger} than
the corresponding EOB prediction. The fact that the $E$-method yields, at 3PN, significantly larger
values of  $({\Omega M_{\rm irr}})_{\rm LSO}$ than the EOB one, in the equal-mass case, is
not surprising in view of the fact that it already did so in the ``test mass limit''
($\nu \to 0$), i.e. when one mass is much smaller than the other one. Let us
introduce the dimensionless ratios,
$ a=  (\Omega M_{\rm irr})/ {(\Omega M_{\rm irr})}^{\rm exact},
b = (E_{\rm real}/M_{\rm irr}-1)/{(E_{\rm real}/M_{\rm irr}-1)}^{\rm exact}$,
for LSO characteristics. In the test-mass limit (where the exact answers are known),
the EOB method yields {\it exact} results,
i.e. $(a,b) = (1,1)$ at {\it all} PN levels. By contrast, the $E$-method yields
\cite{DIS98,DJS00}: $(8, 2.914)$ at 1PN, $(1.824, 1.315)$ at 2PN,
and $(1.275, 1.103)$ at 3PN. This  monotone approach, from upwards, towards the exact
answer is linked to the signs of the expansion coefficients of the function $E(x)$.
As these signs remain the same for all values of $\nu$ one also expects the $E$-method
to yield {\it overestimates} of $\Omega$ and $E_{\rm real}$ in the equal-mass case.
It is, in our opinion, an important feature of the EOB method (which is shared
by none of the non-resummed PN approaches) that it yields exact results when $\nu \to 0$,
and ``robust'' results (not depending very much on the PN accuracy used) when $\nu$ is
not zero, and notably when $\nu = 1/4$ (equal-mass case). On the other hand, in the
equal-mass {\it corotating} case, the 3PN-level  $E$-method predicts a maximal binding
configuration  which is within 10 \% (both for $\Omega$ and  $E_{\rm real}$) of the 3PN
EOB prediction, as well as of the HKV numerical results \cite{B02}.
In view of the test-mass-limit behavior,
and of the rather large difference in the estimates of the effects linked to corotation
(see the distance between the diamonds in Fig.~\ref{f:comp_lso}, and the discussion
in the next subsection), we view this 10 \% agreement as accidental.

Though the 1PN, 2PN and 3PN results
are much closer in the EOB approach than in other (less resummed) PN approaches
(compare, e.g., the results here with Fig. 3 of Ref. \cite{B02} whose 1PN predictions would
not fit on our Figure~\ref{f:comp_lso} above), one, however, notices
that there is somewhat of a jump between the 1PN and 2PN results on one hand, and the
3PN ones on the other hand. This is due to the largish positive numerical value of
$a_4^{\rm GR}$, Eq. (\ref{eq2.11}). Indeed, a positive $a_4$ means a repulsive term
in the radial Hamiltonian. This (relative) ``repulsion''
allows the LSO to move inwards, towards more binding.
The largish positive value of $a_4^{\rm GR}$ (when $\nu = 1/4$) has thereby a
somewhat significant effect on the maximal binding (towards more binding).
One should, however, remember that Ref. \cite{BD00} has shown
that the location of the LSO was blurred by radiation damping effects. The difference
between 2PN and 3PN predictions is probably barely observable in the GW signal
emitted by coalescing black holes. [See \cite{DIJS} for a discussion of the observability
of the various parameters entering the EOB framework, and for an estimate of the
plausible value of the 4PN term $a_5^{\rm GR}$ and its effect on the LSO.]

 But the most striking feature
of Fig.~\ref{f:comp_lso} is, on the one hand, the closeness between the EOB results and the numerical
results of \cite{GGB2}, and on the other hand, the huge difference between the latter
results, and the numerical results of the conformal imaging \cite{cook,PfeifTC00}
or puncture \cite{baumgarte} IVP methods.

\begin{figure}
\includegraphics[height=8cm]{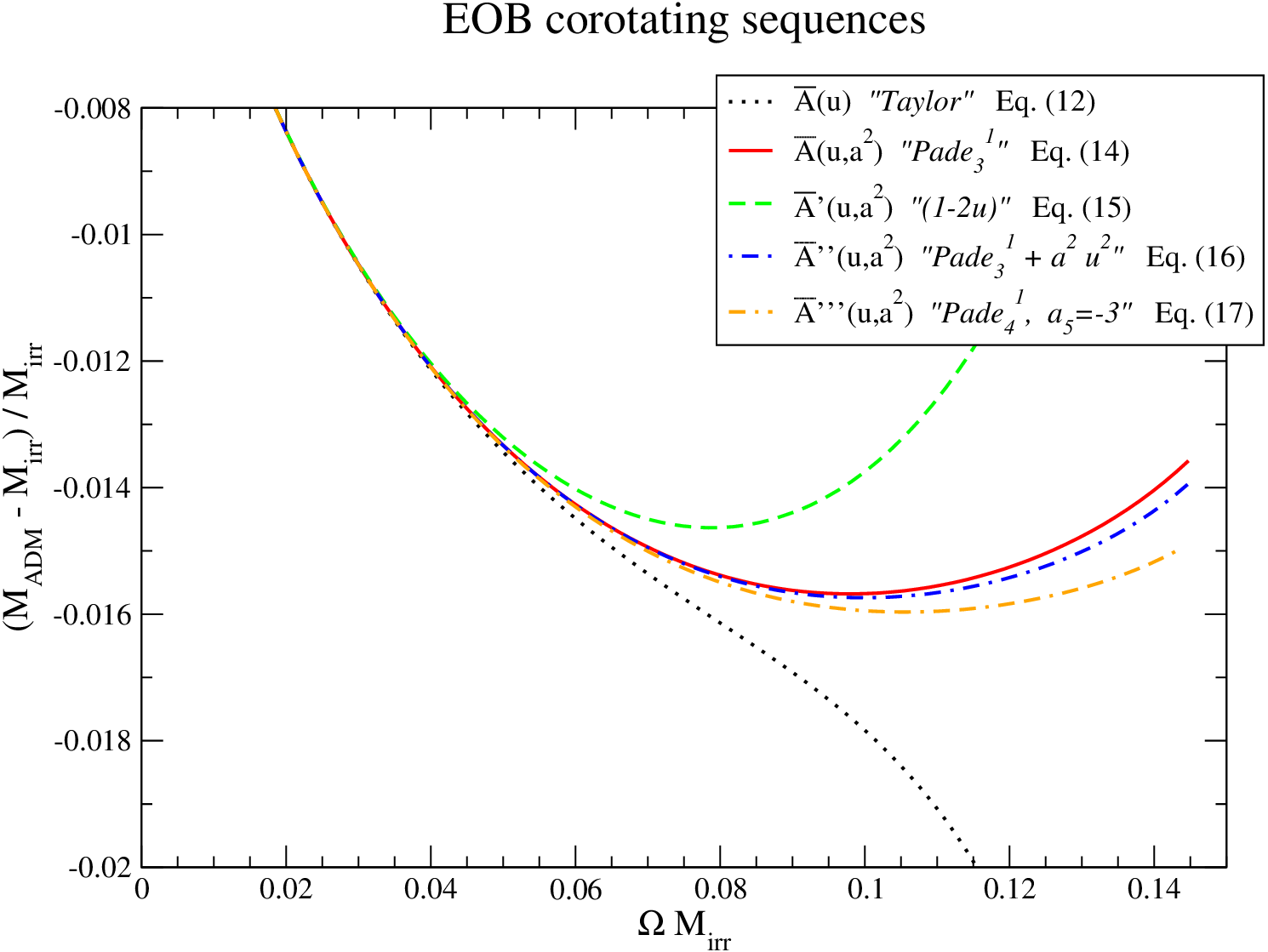}
\caption{\label{f:seq_abar} Binding energy as a function of the orbital velocity,
for the various alternatives to the $\overline{A}(u,{\hat a}^2)$ function.}
\end{figure}

To complete the information displayed in Fig.~\ref{f:comp_lso} we give in
Table~\ref{t:lso}
the corresponding numerical data. Table~\ref{t:lso} gives also data corresponding to several
possible alternative versions of the basic function $\overline{A} (u)$ used in the EOB approach
(see Eqs.(\ref{eq2.15})-(\ref{eq2.17})). In particular, note that all the Pad\'e alternatives
($\overline{A}, \overline{A}'', \overline{A}'''$) yield very close results.
This is further illustrated in Fig.~\ref{f:seq_abar}.
Even the ``factorized Taylor'' version ($\overline{A}'$) (which incorporates a rather brutal way
of forcing $\overline{A}(u)$ to have a simple zero) gives (as is clear from Fig.~\ref{f:comp_lso})
results which are in good agreement both with other 3PN EOB results and with the HKV ones.
Though we believe that such a ``factorized Taylor'' version of $\overline{A}(u)$ is, a priori,
not as good as the various Pad\'e versions, we included it because we think that its
difference with the Pad\'e results gives a plausible upper limit of the ``error bar''
on the 3PN EOB predictions. For instance, we think that a plausible range for the correct
maximal binding energy is $ E^{\rm corot}_{\rm real}/M_{\rm irr} -1 = -0.0157 \pm 0.0011$,
in the corotating case,
and $ E^{\rm irrot}_{\rm real}/M_{\rm irr} -1 = -0.0167 \pm 0.0009$ in the irrotational one.
Note also that Table~\ref{t:lso} (and Fig.~\ref{f:comp_lso}) include no data corresponding
 to the straightforward Taylor
version (\ref{eq2.12}) [truncated at ${\cal O} (u^5)$]. Indeed,  the
qualitatively different behavior of $A^{\rm Taylor} (u)$ (without simple zero, see
Fig.~\ref{f:A(u)-var} above) prevents the presence of a minimum along the curve $E(\Omega)$,
as shown in Fig.~\ref{f:seq_abar}. The latter Figure illustrates the fact that, even within the
EOB approach, there is a contrast between the closeness of the Pad\'e-type predictions and the
dispersion of the Taylor-type ones (straightforward Taylor and factorized Taylor).
 Again, we can use the distance (visually displayed in Fig.~\ref{f:seq_abar})
between Pad\'e results and Taylor ones to define a (two-sided) ``error bar'' on the EOB
Pad\'e predictions.


\begin{table*}
\caption{\label{t:lso}Parameters of the LSO configuration according to various
methods. Unless otherwise noticed, PN computations correspond to the
EOB approach with $\overline{A} (u)$
functions given in the text and with the 3PN parameter $a_4$ set to $4.672$,
see Eq.~(\ref{eq2.11}). For $\overline{A}''' (u , \widehat{a}^2)$,
we have used $a_5=-3$. The relative ``orbital'' binding energy $e$ is defined by
Eq.~(\ref{e:e_def}).}
\begin{ruledtabular}
\begin{tabular}{lllllllll}
Method & $\Omega M_{\rm ir}$ & $M_{\rm ADM}/M_{\rm irr}-1$ & $J/M_{\rm irr}^2$ &
$ L/M_{\rm irr}^2$ &  $u$  & $\hat a$ &  $m_1/m_1^{\rm irr}$ & $e$ \\
\hline
1PN corot  & 0.0667 &  -0.0133 &  0.907 &  0.846 &  0.164 &  0.108 &  1.0019
		&  -0.0152 \\
2PN corot  & 0.0715 &  -0.0138 &  0.893 &  0.828 &  0.173 &  0.115 &  1.0022
		& -0.0160 \\
3PN corot $\overline{A} (u , \widehat{a}^2)$
           & 0.0979 &  -0.0157 &  0.860 &  0.774 &  0.220 &  0.149 &  1.0037
           	&  -0.0193 \\
3PN corot $\overline{A}' (u , \widehat{a}^2)$
          & 0.0789 &  -0.0146 &  0.876 &  0.805 &  0.186 &  0.125 &  1.0026
          	& -0.0172 \\
3PN corot $\overline{A}'' (u , \widehat{a}^2)$
          & 0.0999 &  -0.0157 &  0.859 &  0.773 &  0.223 &  0.151 &  1.0037
          	& -0.0193 \\
3PN corot $\overline{A}''' (u , \widehat{a}^2)$
          & 0.1055 &  -0.0160 &  0.856 &  0.766 &  0.233 &  0.157 &  1.0041
          	& -0.0200 \\
3PN corot non resum. \cite{B02}
	  & 0.091  &  -0.0153 &	       &       &        &        &    &     \\
HKV corot \cite{GGB2} & 0.103  &  -0.017  &  0.839 &       &     &     &      &         \\
\hline
1PN irrot \cite{BD00} & 0.0692 &  -0.0144 &  0.866 &  0.866 &  0.167 & 0 & 1 & -0.0144 \\
2PN irrot \cite{BD00} & 0.0732 &  -0.0150 &  0.852 &  0.852 &  0.174 & 0 & 1 & -0.0150 \\
3PN irrot $\overline{A} (u , 0) = \overline{A}'' (u , 0)$ \cite{DJS00}
          & 0.0882 &  -0.0167 &  0.820 &  0.820 &  0.202 & 0 & 1 & -0.0167 \\
3PN irrot $\overline{A}' (u , 0)$
          &  0.0786 & -0.0158 & 0.834  & 0.834  & 0.185  & 0 & 1 & -0.0158 \\
3PN irrot $\overline{A}''' (u , 0)$
          & 0.0898  & -0.0168 & 0.817  & 0.817  & 0.205  & 0 & 1 &  -0.0168 \\
3PN irrot non resum. \cite{B02}
	  & 0.129  &  -0.0193 &	 0.786 &       &        &        &    &     \\
IVP-punct irrot \cite{baumgarte}
          & 0.18   &  -0.023 &  0.737  &      &       &  &  &  \\
IVP-conf irrot \cite{PfeifTC00}
          & 0.166  & -0.0225 & 0.744   &       &       &  &  &
\end{tabular}
\end{ruledtabular}
\end{table*}

\subsection{Spin effects in the determination of the LSO}

Before considering in more detail the characteristics of the circular orbits around
the LSO, it is interesting to discuss the reason why there is such little difference
between the irrotational and corotating cases. Indeed, if we consider, say,
our preferred EOB potential $\overline{A} (u, {\hat a}^2)$, at the 3PN approximation,
Table~\ref{t:lso} shows that the maximum corotating binding energy is $- 0.0157$, while the
irrotational value is $- 0.0167$. The difference is only $ 1 \times 10^{-3}$. This
difference is about 4 times smaller than the difference obtained in  \cite{B02}
which used a non-resummed PN approach. We are going to see that the
relatively large difference obtained in this non-resummed PN approach is rooted in the
(truncated) {\it perturbative} nature of this approach.

 One can distinguish two leading contributions
to the difference in binding energy: (i) a contribution linked to the (kinetic)
rotational energies of the spinning black holes, and (ii) a contribution linked to the
spin-orbit interaction. [There are also mixed terms involving, e.g., products of the
spin kinetic energy with orbital effects, but one can check that they are numerically
smaller.] In the language of the present paper, the first contribution is essentially
embodied in the factor  $m_1/m_1^{\rm irr}$ in the first Eq.(\ref{eq3.18})
above. For corotating configurations, Eq.(\ref{eq3.16}) above approximately yields
$\widehat{a}_1 = \widehat{a}_2 = 2 \omega$ where $\omega \equiv M_{\rm irr} \Omega \approx
M \Omega$. This yields  $m_1/m_1^{\rm irr} - 1 \approx \omega^2/2$. Numerically, we have
(for the irrotational 3PN case; see Table~\ref{t:lso}), $\omega \approx 0.09$, so that
 $m_1/m_1^{\rm irr} - 1 \approx 4 \times 10^{-3}$. This is essentially the difference
``corotating - irrotational'' in binding energy obtained in \cite{B02}, because,
within the non-resummed PN approach of \cite{B02}, the kinetic-spin contribution
largely dominates over the spin-orbit interaction energy. We wish, however, to
point out that there are consistency problems in the application of straightforward
(truncated) PN-expanded methods to the determination of the effect of the spin-orbit
interaction on the maximum binding energy. By contrast, we shall see that the EOB estimate of the
same quantity has no consistency problems, and happens to be much larger and  to nearly cancel
the kinetic-spin contribution.

To have a common language between the two approaches (straightforward PN approach
and EOB approach), let us write the Hamiltonian of a binary system as
$H = H_0 (r,L) + H_1 (r,L,S)$, where $H_1 = 2 \, LS / r^3$ is the spin-orbit interaction
to linearized order \cite{BOC}. Here, $S = \sigma_1 S_1 + \sigma_2 S_2$ denotes the effective
spin (with the same notation as above for $\sigma_1 , S_1$,etc.). We only consider
the aligned case where the spins are parallel to the orbital angular momentum $L$.
[Here, we consider the masses as given parameters, because we treat separately
the effects linked to the kinetic energies of the spins.] To first order in $H_1$,
the energy along sequences of circular orbits, as a function of the radial coordinate $r$,
 (obtained by solving $\partial H/ \partial r = 0$ so as to get $ L = L(r)$), can be
 written as $H^{(r)} [r] = H_0^{(r)} [r] + H_1^{(r)} [r]$, where $H_0^{(r)} [r]$ is
 the answer corresponding to $H_0 (r,L)$ and where the first-order correction linked
 to $H_1(r,L)$ ( we consider $S$ as a spectator variable which will be, at the end,
 replaced by a function of $r$ obtained from the corotation condition) reads:
\begin{equation}
H_1^{(r)} [r] = \left[ H_1 (r,L) - \frac{\partial_L \, H_0 (r,L)}{\partial_{r
L}^2 \, H_0 (r,L)} \, \partial_r \, H_1 (r,L) \right]_{L = L_0 (r)}  \ .
\end{equation}
Here, $L_0 (r)$ is the angular momentum along the radial sequences of circular orbits
defined by $H_0 (r,L)$. If one then asks (as in \cite{KWWR,B02}) for the variation
of the energy as a function of the orbital frequency
$\Omega (r) = [\partial \, H (r,L) / \partial \, L]_{L = L(r)}$, one finds
$ H^{(\Omega)} [\Omega] = H_0^{(\Omega)} [\Omega] + H_1^{(\Omega)} [\Omega]$,
where $H_0^{(\Omega)} [\Omega]$ is the answer corresponding to $H_0$ and where the
$H_1$-related correction reads:
\begin{equation}
H_1^{(\Omega)} [\Omega] = \left[ H_1^{(r)} - \frac{d \, H_0^{(r)} / dr}{d \,
\Omega_0 (r) / dr} \left( \partial_L \, H_1 - \frac{\partial_L^2 \,
H_0}{\partial_{r L}^2 \, H_0} \, \partial_r \, H_1 \right)\right] \ .
\end{equation}

Here, the brackets mean that one evaluates a quantity for $L = L_0 (r)$ and then
changes variables via: $r = r_0(\Omega)$.
Having these results in hand, we can now compare the straightforward (non-resummed)
PN approach to the EOB one. The non-resummed PN approach considers that $H_1$ is
already of high-PN order, so that, in all the $H_1$-related corrections above
one can replace the zeroth-order Hamiltonian $H_0$ by its {\it Newtonian}
approximation. Applying then the formulas above to the spin-orbit Hamiltonian
($H_1 \propto L r^{-3}$) yields the simple results:
$H_1^{(r)} = - \frac{1}{2} \, [H_1]$ and $H_1^{(\Omega)} = - \frac{2}{3} \, [H_1]$,
where the brackets indicate that $H_1$ should be evaluated along the considered
sequence (and expressed in terms of the corresponding variable). [One easily checks
that the result for $H_1^{(\Omega)}$ is equivalent to the formula derived in \cite{KWWR}
and used in \cite{B02}.]

It is interesting to note that the results are numerically different. To have
such a difference is fine far from the LSO (and the PN calculations are both valid there),
but it is problematic near the LSO. Let us first recall the standard approximate estimate
(valid to linear order in the perturbation) of the minimum of
 a function of the type $f(x) = f_0(x) + f_1(x)$ where  $f_0(x)$
has a local minimum at $x_{\min}$: when working to first order in $f_1$ one
easily gets $f_ {\min} \approx f_0(x_{\min}) + f_1(x_{\min})$.
Applying this general result to the case at hand, we simply get
the spin-orbit (SO) contributions at maximum binding by evaluating
the corrections  $H_1^{(r)}$ or $H_1^{(\Omega)}$ at the zeroth-order (irrotational) LSO.
We thereby get different values (by a factor $4/3$).
However, it is physically (and mathematically) inconsistent to get different estimates
of the SO-induced change in maximum binding energy. Indeed, one is considering the
same quantity (the energy) expressed in terms of different
(gauge-invariant\footnote{We recall that the EOB (Schwarzschild-like) radial coordinate is
 defined in a gauge-invariant way.}, monotonically varying)
independent variables: $r$ or $\Omega$. The numerical value of the minimum of this
quantity should be the same (even if is reached at a different value of the independent
variable).
This consistency problem is linked to the perturbative nature of the (non-resummed) PN approach
which (self-consistently) treats the spin-orbit effects as being of higher PN order,
and truncates away (most of) the ``mixed'' terms involving products of the PN contributions
to $H_0$ with SO effects. Note, a contrario, that the consistency of the non-resummed PN
treatment would be recovered if one were to keep all those mixed terms.

 By contrast, the EOB approach is immune to such consistency problems.
 Indeed, the EOB approach does not approximate $H_0$ in the
correction terms by its Newtonian expression. It, instead, uses the full (non-perturbative)
expression Eq.(\ref{eq2.13}). This value of  $H_0$ consistently contains an LSO, i.e.
 a value of $r$ (and $\Omega$) where $d H_0^{(r)} / dr$ vanishes. One then sees that,
 when evaluating (in the approximate way indicated above) the change in the maximum
 binding induced by $H_1$, the last correction term in $H_1^{(\Omega)}$ vanishes,
 so that we get (whatever be the independent variable used)
$\delta E_{\rm LSO}^{\rm EOB} = {H_1^{(r)}}_{\rm LSO} = {H_1^{(\Omega)}}_{\rm LSO}$.
Moreover, using the full expression  Eq.(\ref{eq2.13}) for $H_0$ one gets the
explicit expression (where $u = M/r$):
\begin{equation}
\delta \, E_{\rm LSO}^{\rm EOB} = \{ 1 - 3 \, A / (2 \, A + u \, \partial_u \,
A)\} \, [H_1]  \ .
\end{equation}
This value depends on the expression used for the EOB potential $A(u)$, and, thereby,
on the corresponding location of the LSO.
If we consider, for simplicity, the limit $ \nu \to 0$, we can use the simple
form $A(u) = 1 - 2u$, and the corresponding LSO location: $u_{\rm LSO} = 1/6$.
This then yields $\delta E_{\rm LSO}^{\rm EOB} = - [H_1]$. We see that the EOB
approach gives a larger contribution than both the straightforward PN methods
discussed above. This larger value is due to the fact that the EOB approach is
{\it non perturbative} in that it treats the effective potential $A(u)$ exactly
near the LSO.

  Let us introduce the dimensionless ``orbital'' binding energy
\begin{equation}  \label{e:e_def}
	e \equiv \frac{E - M}{M}
		= \frac{1}{4} \widehat E - 1
		= \frac{1}{4} \frac{E_{\rm real}}{\mu} - 1
		= \frac{m_1^{\rm irr}}{m_1} \frac{M_{\rm ADM}}{M_{\rm irr}} - 1 \ .
\end{equation}
Inserting the explicit value of the spin-orbit Hamiltonian
(and consistently using $r_{\rm LSO} = 6 M$) finally gives the following estimate
for the spin-orbit modification of $e$, in the ``test-mass'' limit $ \nu \to 0$:
\begin{equation}
\delta^{\rm SO} \, e_{\rm LSO} = - \sqrt 3 \ 6^{-3} \, \nu_4 \, \widehat a =
-0.802 \times 10^{-2} \, \nu_4 \, \widehat a  \ .
\end{equation}
Here, $\nu_4 \equiv 4 \nu$ and $\widehat a$ denotes the effective dimensionless
spin parameter $S/M^2$. The numerical coefficient $0.802$ is about twice larger
than the corresponding result obtained in \cite{B02}. Note that this coefficient
is consistent with the small $\nu$ limit of the result written in Eq.(4.1) of \cite{D01}.
The latter equation then shows that, in the equal-mass case (of interest here), $\nu = 1/4$,
i.e. $\nu_4 = 1$, the numerical coefficient giving the SO-induced change of $e_{\rm LSO}$
is further enhanced (by a factor 1.888) to the value
$\delta^{\rm SO} \, e_{\rm LSO} = -1.52 \times 10^{-2} \widehat a$. 
Actually, the data in Table~II of \cite{D01} show that the latter result is valid
only for  values of $\widehat a$ smaller than $0.1$. For the value $\widehat a \approx 0.15$
relevant to us (see Table~\ref{t:lso} above), nonlinear effects in $\widehat a$ further enhance
the change in binding energy. Interpolating between the numerical data given in
Table~II
of \cite{D01} one can see that a better estimate of $e$ for $\widehat a \approx 0.15$
is $e_{\rm LSO} \approx - 0.0197$. Compared to the value without spin-orbit contribution,
$e_{\rm LSO} = - 0.0167$, this yields a SO-related modification equal to
$\delta^{\rm SO} \, e \simeq - 0.30 \times 10^{-2}$.

Finally, we conclude that the EOB approach gives (at 3PN): (i) an increase of
$E_{\rm LSO}/M_{\rm irr}$ 
due to the kinetic energy of the spinning black holes of order $ + 4 \times 10^{-3}$,
and a decrease linked to spin-orbit effects (as estimated within
the EOB framework) of order $ - 3 \times 10^{-3}$. The net result is that
 $E_{\rm LSO}/M_{\rm irr}$ increases only by  $ + 1 \times 10^{-3}$. 
This analysis has clarified why and how the EOB approach yields a rather small
final effect for the difference between the irrotational case and the corotating one.
We view this result as a confirmation of the generic robustness of the EOB approach.
By contrast, we view the significantly larger change predicted by non-resummed
PN approaches as
further confirmations of the generic  lack of robustness of non-resummed PN
methods. An interesting lesson of this comparison is that the larger SO-related changes
predicted by the EOB approach are crucially linked to its {\it non perturbative} nature.

\begin{figure}
\includegraphics[height=8cm]{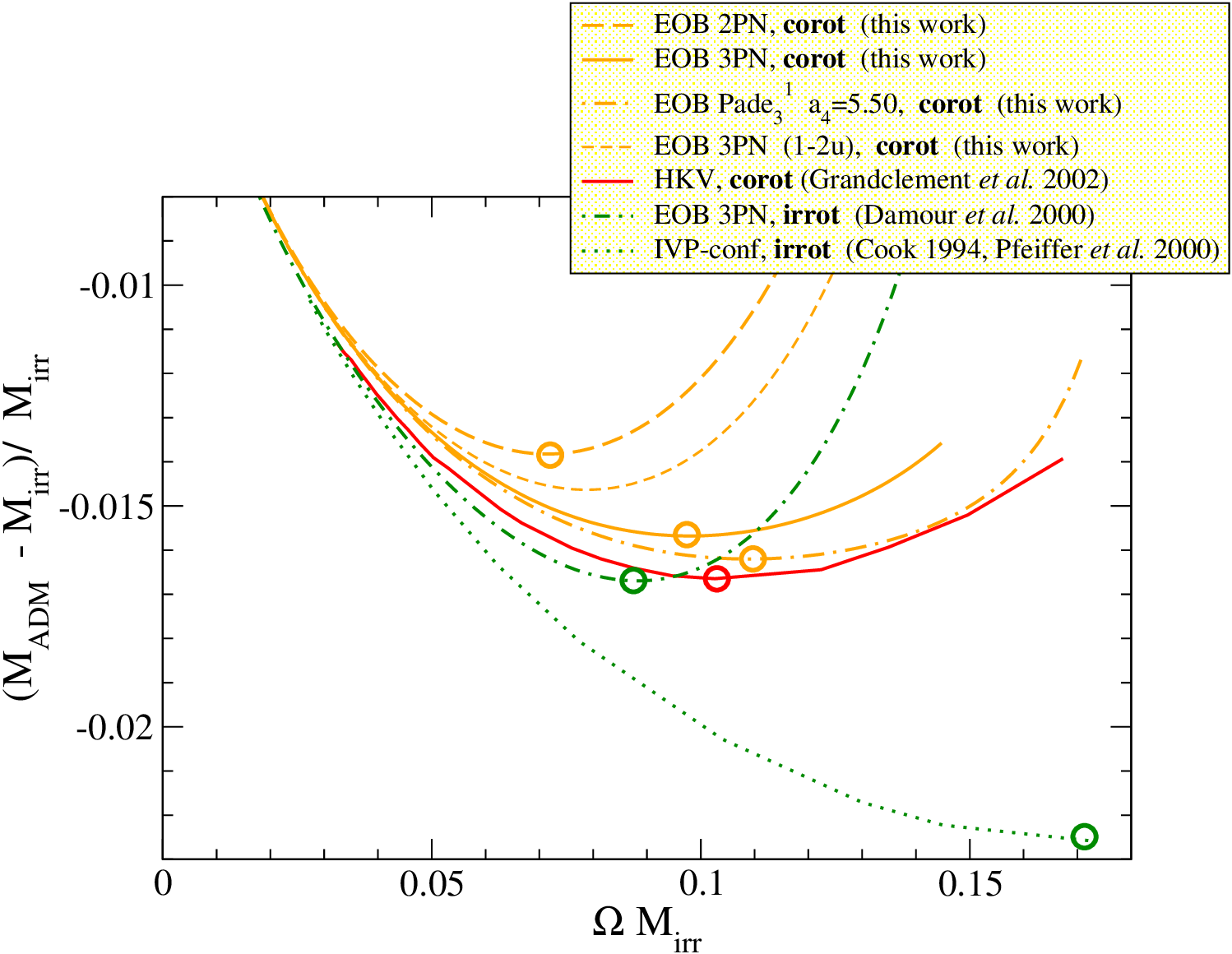}
\caption{\label{f:seq_e} Binding energy as a function of the orbital
angular velocity, according to various analytical and numerical methods.}
\end{figure}

\begin{figure}
\includegraphics[height=8cm]{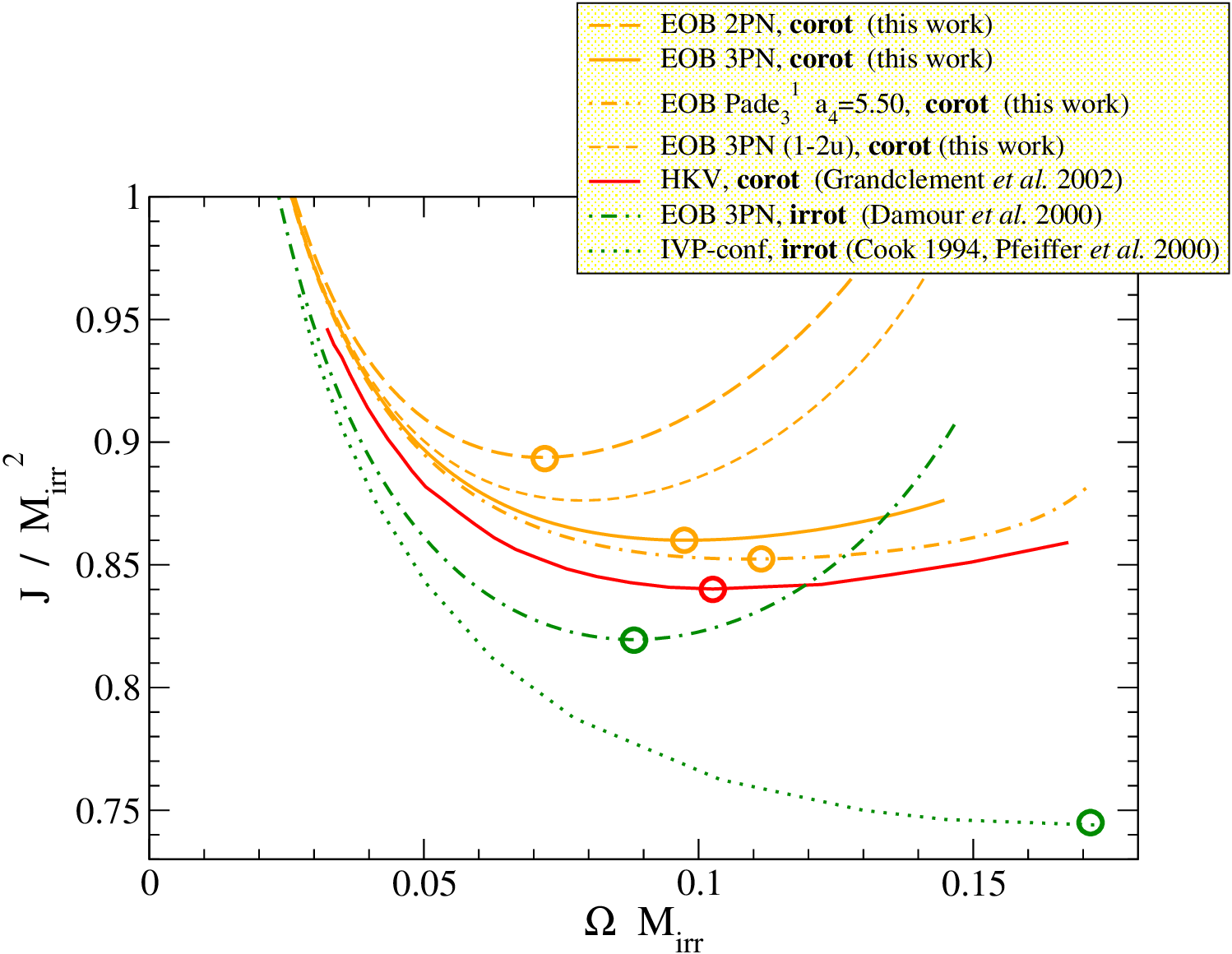}
\caption{\label{f:seq_j} Total angular momentum as a function of the orbital
angular velocity, according to various analytical and numerical methods.}
\end{figure}

\subsection{Evolutionary sequences}

As we said above, the LSO characteristics do not embody the really useful information 
contained in the various approximations to binary black holes dynamics. To have a complete 
hold on the two-body dynamics one would need to compare the Hamiltonians, $H(r,p_r , L , 
S_a)$. The EOB method provides such a complete description (which allows it to describe not 
only the adiabatic inspiral but also the crucial transition to the plunge). However,
numerical methods do not give access to such a multi-variable function. As a makeshift we 
can at least compare various functions of one variable linked to the two-body dynamics. In 
Fig.~\ref{f:seq_e} we plot various analytical and numerical estimates of the binding energy
of circular orbits versus the orbital frequency, while Fig.~\ref{f:seq_j}
plots the total angular momentum $J$.
These figures show that the methods that led to good agreement for LSO characteristics in 
Fig.~\ref{f:comp_lso}, also lead to good agreement for all circular orbits, and all
available dynamical 
quantities. We have also included in these figures the results coming out of the 
straightforward truncation of the $A$-function, Eq.~(\ref{eq2.12}). We see that it agrees
rather well (compared, say, with the conformal-imaging or puncture data) with the helical
Killing vector (HKV) data nearly up to the LSO.

If we remember that Ref.~\cite{BD00} has shown that, a little bit above the normal
(adiabatic) LSO, radiation damping effects dominate over the ``restoring'' radial potential, 
it is very plausible that even the pure Taylor EOB function $\overline A (u)$ would lead to 
a GW signal essentially indistinguishable from the one obtained from a Pad\'e-resummed 
function $\overline A (u)$. Note that Figs.~\ref{f:seq_e} and \ref{f:seq_j}
confirm the messages of Fig.~\ref{f:comp_lso}: the main 
feature is a nice agreement of all versions of EOB and HKV results versus a  significant
difference with the conformal-imaging or puncture results. When looking at the finer 
structure of the EOB predictions one sees that the 3PN predictions are much closer to the 
HKV data than the 2PN ones.

 As we said above, the multi-parameter ``flexibility'' of the EOB 
approach $(a_4 (\nu) , a_5 (\nu) , a_6 (\nu) , \ldots)$ can be exploited, to further 
improve, if deemed necessary, the agreement with numerical data. For instance, we have 
played with the addition of $a_5 (\nu)$ and found that, with the Pad\'e resummation 
(\ref{eq2.17}), a value $a_5 \left(\frac{1}{4} \right) \simeq \, -3$ further improves the 
agreement with HKV data. However, we do not consider such a best fit value of $a_5 (\nu)$ as 
significant at this stage. Another interesting ``robustness'' exercise consists in modifying 
the general relativistic (GR) prediction (\ref{eq2.10}) for the 3PN coefficient 
(\ref{eq2.10}). We recall that, from the recent dimensional continuation calculation of 
\cite{DJS01}, the GR value should be, when $\nu = \frac{1}{4}$, $a_4^{\rm GR} = 4.672$ (see 
Eq.~(\ref{eq2.11})). By varying $a_4$ (without adding any further terms $a_5 (\nu) ,
 a_6 (\nu), \ldots$) we found that a value which provides a slightly better 
fit is $a_4 \left(\frac{1}{4} \right) = 5.50$. The corresponding curves are plotted in 
Figs.~\ref{f:seq_e} and \ref{f:seq_j}.
On the other hand, note that the value $a_4 = 0$, corresponding
to the 2PN approximation only, noticeably worsens the agreement with the HKV data.
In all cases, the 
reader should notice that the binding energy differences linked to these different choices 
for the 3PN contribution $a_4$ remain small compared with the dispersion among all numerical 
results.

\section{Conclusions}\label{sec4}

\subsection{Robustness of the EOB method}

The main conclusions of this work are the following. We have confirmed the robustness of the
effective one-body (EOB) approach to binary black hole dynamics. The various ways of
resumming the crucial EOB radial function $\overline A (u)$ exhibited in
Figs.~\ref{f:A(u)-var} and \ref{f:A(u)-Pade}
lead to predictions for the binding energy and the angular momentum as functions of orbital
frequency, which are very close to each other, see Figs.~\ref{f:comp_lso},
\ref{f:seq_e} and \ref{f:seq_j}, and Table~\ref{t:lso}. Even the ``non
resummed'' function $\overline A (u)$ (upper curve in Fig.~\ref{f:A(u)-var})
leads to dynamical predictions
which are quantitatively close to the ``resummed'' predictions up to the last orbits before
the Last Stable (circular) Orbit (LSO) (see Figs.~\ref{f:seq_e} and \ref{f:seq_j}).
We have argued that it would
probably yield indistinguishable waveforms if used, as in \cite{BD00}, to study the
transition between inspiral and plunge. [However, if one wishes, as in \cite{BD00}, to
continue the description of the plunge nearly down to coalescence we expect that a non
resummed $\overline A (u)$ will quickly lead to problems and to strong deviations from the
results obtained from various resummed $\overline A (u)$.] Another robustness feature of the
EOB approach is its good post-Newtonian (PN) convergence properties. This was illustrated in
Figs.~\ref{f:comp_lso}, \ref{f:seq_e} and \ref{f:seq_j} where the 1PN, 2PN and 3PN predictions
are seen to be all much closer to
the helical Killing vector (HKV) data than to the other numerical data. Note, however,
that the best agreement with HKV data is obtained for the 3PN prediction. We emphasized
also the consistency (and robustness) of the EOB approach in the treatment of the
changes induced by spin effects. This contrasts with non-resummed PN approaches
which predict different changes when one uses different parameters to run along a
corotating sequence.

\subsection{Extension of the EOB method to corotating systems}

The most novel aspect of the present work is that we have shown how to apply to corotating
sequences the recently defined \cite{D01} extension of the EOB approach to spinning black
holes. The effects linked to the spins turn out, finally, to be rather small. For instance,
they are significantly smaller than the effect of using the 3PN-accurate EOB Hamiltonian
instead of a 2PN-accurate one, see Fig.~\ref{f:comp_lso}. We discussed the reasons behind
this behavior (essentially a near cancellation between an energy increase linked to
the spin kinetic energy and an energy decrease linked to the rather strong spin-orbit
interaction predicted by the EOB method).

\subsection{Good agreement between EOB and HKV results}

Our findings fully confirm what was announced in \cite{GGB2}: the new HKV numerical data are
much closer to analytical results than previous numerical data. At face value, this is a
very encouraging fact which suggests that one can now implement the philosophy which was at
the basis of the EOB approach \cite{BD00}: to use a (resummed) analytical approach to
describe binary black holes not only during early inspiral (which was previously thought to
be the only possible domain of validity for an analytical approach), but also during late
inspiral and, most importantly, during the transition between inspiral and plunge. Numerical
relativity techniques can then be used only for describing the rest of the plunge and the
coalescence. [As noted in \cite{baker} one might also simplify matters by describing the end
of the coalescence by the ``close limit approximation''.] A crucial requirement for
successfully merging together analytical and numerical results is to be able to match the
analytical description just after the last stable orbit (LSO) with suitable numerical
initial data. This paper is an encouraging step in this direction, in view of the robust and
close agreement between EOB results and HKV ones.

However, this agreement raises many questions. A first question is linked to the
``approximations'' used in the present implementation of the HKV philosophy. Indeed, though
this method is not, in principle, limited to conformally flat data, its present
implementation has, for simplicity, imposed conformal flatness of the spatial metric, and
has reduced the ten equations  (for ten unknowns) to solve to a subset of five equations
(for five unknowns). The reason why this restriction is problematic is that, from the
analytical side, the values of the successive coefficients in the various functions entering
the EOB formalism depend quite sensitively on the exact form of the metric, and, in
particular, on the fact that it is not conformally flat. For instance, it was shown in
\cite{DJS00} that both the ``effective potential'' $A(u)$, and the energy map (\ref{eq2.1}),
are modified, at the 2PN approximation, if one truncates Einstein equations to  impose
conformal flatness. The corresponding modification at the 3PN approximation is currently not
known. However, these modifications might turn out not to be critical in view of the
robustness of the EOB predictions. Fig.~\ref{f:comp_lso}
shows that even if we use the EOB Hamiltonian at
the 1PN approximation (which, in fact, essentially coincides with the ``Newtonian''
approximation, i.e. the EOB reformulation of the ``test-mass approximation''), one is much
closer to HKV data than to the other numerical data.

\subsection{Plausible explanation of the disagreement with IVP results}

The most important question
posed by the agreement analytical/HKV is to understand why the IVP approaches give
(consistently among themselves) drastically different results. After all, both types
of numerical approaches
use (at present) the same simplification of spatial conformal flatness $(\gamma_{ij} =
\psi^4 \, \delta_{ij})$, and they both numerically construct solutions to the Hamiltonian
and momentum constraints (approximate solution to the momentum constraint for HKV).
Seen from this technical point of view, the difference between the
two schemes lies in the way of solving the momentum constraint.
In the IVP method~\cite{cook,PfeifTC00,baumgarte}, it is
solved by means of a simple ansatz
(Bowen-York extrinsic curvature \cite{BowenY80} or its isometric version
\cite{KulkaSY83})
for the second fundamental form $K_{ij}$ whose
weak-field limit is reasonable, while in the HKV method~\cite{GGB1,GGB2} one
determines an essentially unique
solution for $K_{ij}$ by the requirement that it be induced by restricting a
helically-symmetric spacetime to a Cauchy hypersurface
(see also \cite{Cook02} for a discussion).
We think that the basic physical
difference between the two approaches, which underlies their drastically different
predictions, is the following. Though it must be admitted that the imposition of a
helical symmetry can only be an approximation, it is a physically well motivated
approximation which has the great virtue of selecting a specific solution of the
constraints. This is very similar to what happens in the post-Newtonian schemes, especially
in their ADM version. Indeed, in the Hamiltonian ADM approach one uses (when dealing with
the near-zone field) the post-Newtonian ansatz $(c^{-1} \, \partial_t \, h \ll \partial_x \,
h)$ to derive and solve an iterative set of coupled {\it elliptic} equations for
$\gamma_{ij}$ and $K_{ij}$. This selects a specific solution of the constraints.
Intuitively, one can think of this (essentially) unique solution selected by the PN method
as the only solution of the constraints which contains the ``right'' amount of free
gravitational wave (GW) data $(h_{ij}^{TT} , \pi_{TT}^{ij})$ which has been generated by the
many previous orbits of the system, and which has therefore reached a quasi-stationary
state. Any other amount of GW data would be a priori allowed at a specific moment of time
but would not correspond to a quasi-stationary state, and would therefore be expected to
``fly away'' with the velocity of light, i.e. to evolve in a violently non quasi-stationary
way, with $c^{-1} \, \partial_t \, h \sim \partial_x \, h$ instead of $c^{-1} \, \partial_t
\, h \ll \partial_x \, h$. We think that this PN understanding of the selection, by
reduction to elliptic equations, of a specific ``quasi-stationary'' solution applies, {\it
mutatis mutandis}, to the HKV approach. The imposition of a helical Killing vector
selects the only ``quasi-stationary'' solution corresponding to a steady situation generated
by a slow inspiral, while the arbitrary choice, in the other methods, of a technically
simple solution of the momentum constraint corresponds to a drastically non-steady state,
containing a wrong amount of free GW data\footnote{This wrong amount of GW data
in the IVP results has been confirmed by a recent study \cite{PfeifCT02}.},
which is expected to ``fly away'' as soon as one tries to evolve the system in time.
If this view is correct, it diminishes the signification of the
``effective potential'' approach applied to the initial data of
\cite{cook,PfeifTC00,baumgarte}. Indeed, these data correspond to a non-steady state
to which it is incorrect to apply any energy minimization principle.

\subsection{Fitting the EOB Hamiltonian to numerical data}

We have started to exploit the ``flexibility'' of the EOB approach (already emphasized in
\cite{D01}), i.e. the use of the expansion coefficients appearing in the EOB approach as
parameters to be fitted. For instance, Figs.~\ref{f:seq_e} and \ref{f:seq_j}
exhibit the dynamical predictions
obtained from a modified 3PN coefficient, $a_4^{\rm best \, fit} \left(\frac{1}{4} \right)
\simeq 5.50$ instead of $a_4^{\rm GR} \left(\frac{1}{4} \right) \simeq 4.67$. The modified
value of $a_4$ leads to a better agreement with the HKV data. As we are not adding any
higher PN contributions, $a_5 (\nu) u^5 + a_6 (\nu)u^6 + \ldots$, such a best fit value
for  $a_4$ can be viewed as a way of mimicking the effect of such (missing) higher PN
terms.

We resisted the temptation to
(introduce and) vary more $a_n (\nu)$ parameters up to getting a really close agreement with
HKV data, because, at this stage, such a formal exercise would not be justified. It is,
however, important to keep in mind the suggestion made in \cite{D01} that a suitable
``numerically fitted'' EOB Hamiltonian may be a very useful tool for combining the advantages
of analytical and numerical methods. If we look ahead to the problem of exploring the
dynamics of two arbitrarily spinning black holes, one will have to face an extremely large
parameter space. It seems clear that numerical relativity will not be able (before many
years) to cover densely this parameter space. On the other hand, one can use some sparse
numerical data to determine free parameters introduced in a generalized EOB Hamiltonian.
Then this ``numerically fitted'' EOB Hamiltonian can be used to interpolate between the
sparse numerical data.

\end{document}